\documentclass[
  aps,
  prx,
  reprint,
  floatfix,
  superscriptaddress
]{revtex4-2}

\usepackage{amsmath}
\usepackage{amssymb}
\usepackage{dsfont}
\usepackage{graphicx}
\usepackage{xcolor}
\usepackage{hyperref}
\usepackage{microtype}

\hypersetup{
  colorlinks=true,
  linkcolor=blue!55!black,
  citecolor=blue!55!black,
  urlcolor=blue!55!black
}

\begin{document}

\title{Biorthogonal Conformal Dynamics in Non-Hermitian Quantum Quenches}

\author{Yifan Liu}
\affiliation{Institute for Solid State Physics, University of Tokyo, Kashiwa, Chiba 277-8581, Japan}

\date{\today}

\begin{abstract}
Global quenches at one-dimensional critical points admit a boundary
conformal field theory (BCFT) description in which Euclidean strip
correlators are analytically continued to real time. We formulate this
construction for interacting non-Hermitian critical systems, where a
right ket alone does not specify the dynamical readout and the left
covector is part of the microscopic quench protocol. Independent left
and right preparations flowing to the same conformal boundary define
the two temporal boundaries of a strip with generally complex one-sided
extrapolation parameters; linear and antilinear symmetry pairings are exploited to 
constrain this geometry. In the interacting Yang--Lee spin chain,
statically calibrated boundary data determine the biorthogonal dynamics of the complete-character
return amplitude, a primary one-point function, and a spatial
correlator. The imaginary part of the linear-paired extrapolation
parameter predicts the temporal center of an independently evolved
antilinear-paired one-point function at the $10^{-3}$ relative level;
the same preparation phase controls local phase evolution and the
analytic-continuation path of boundary blocks. Further results test this formalism with a
direct field-on quench, mixed left and right preparations, and a complex
five-state Potts fixed point with complex primary dimensions. These
results establish a BCFT framework for biorthogonal global quenches in
interacting non-Hermitian critical systems, in which complex
temporal-boundary data organize universal post-quench dynamics.
\end{abstract}

\maketitle

\section{Introduction}
\label{sec:introduction}

A global quantum quench probes how equilibrium universality constrains real-time dynamics. In one-dimensional critical systems, the Calabrese--Cardy construction transforms this dynamical problem into a geometric one: a short-range initial state is treated as a regularized conformal boundary state, allowing correlation functions on the resulting Euclidean strip to be analytically continued to real time \cite{CalabreseCardy2006,CalabreseCardy2007,CalabreseCardy2016,Cardy2014,wen2016bridginggloballocalquantum}.
Beyond quenches exactly to criticality, field-theory approaches have
also characterized interacting dynamics in near-critical systems,
including the Ising field theory in a magnetic
field~\cite{DelfinoViti2017,HodsagiKormosTakacs2018}.
This framework yields predictions far beyond simple return probabilities. A primary scaling dimension determines the leading one-point dynamics; conformal blocks with boundary sewing determine multipoint correlators, and a replica geometry determines the growth of entanglement entropy. Recent works further demonstrate that critical Loschmidt amplitudes can extract boundary spectra and local critical exponents \cite{CarignanoTagliacozzo2025,BouComasEtAl2026}. Consequently, a single temporal-boundary geometry maps detailed static conformal field theory (CFT) data onto a hierarchy of nonequilibrium observables.

Non-Hermitian dynamics fundamentally redefines this quench protocol. Non-Hermitian Hamiltonians naturally emerge as effective generators in resonant open systems and conditioned quantum trajectories~\cite{Rotter2009,DalibardCastinMolmer1992,Daley2014,AshidaGongUeda2020}. In this context, pre- and post-selected quantum processes are governed by transition operators constructed from independent right and left states~\cite{AharonovBergmannLebowitz1964,NakataEtAl2021}. To define the post-quench evolution for non-Hermitian systems, we adopt the standard biorthogonal time evolution~\cite{TangKouSun2022,LuChang2026}:
\begin{align}
 |R(t)\rangle
 &=
 \mathrm{e}^{-\mathrm{i}Ht}|R_0\rangle,
 &
 \langle L(t)|
 &=
 \langle L_0|\mathrm{e}^{+\mathrm{i}Ht},
\end{align}
where $H$ is the critical non-Hermitian Hamiltonian and $|L_0\rangle$ and $|R_0\rangle$ are a pair of initial states.
One may also define the normalized biorthogonal transition operator:
\begin{align}
 \rho_{RL}(t)
 =
 \frac{
 |R(t)\rangle\langle L(t)|
 }{
 \langle L(t)|R(t)\rangle
 },
 \qquad
 \mathrm{i}\partial_t\rho_{RL}
 =
 [H,\rho_{RL}].
 \label{eq:intro-biorthogonal-}
\end{align}
Although normalized to unit trace, $\rho_{RL}$ is generally neither Hermitian nor positive semidefinite. Its expectation values represent complex transition amplitudes—conceptually tied to weak values \cite{AharonovAlbertVaidman1988,NakataEtAl2021}—rather than standard probabilistic averages from the  matrix. The left covector therefore explicitly specifies the microscopic transition or readout protocol under investigation, independent of the right preparation.

Consequently, biorthogonal quantum dynamics has emerged as a central paradigm in non-Hermitian quench physics. It provides the foundation for formulating dynamical topology and dynamical quantum phase transitions \cite{QiuEtAl2019,ZhouEtAl2018}, defining non-Hermitian Loschmidt echoes \cite{TangKouSun2022,JingEtAl2024, DoraMoca2024}, and characterizing observables, entanglement, and quantum geometry \cite{LuShiSun2025,LuChang2026}. Correlations and information spreading have been extensively investigated across exactly solvable non-Hermitian spin chains, $PT$-symmetric and quenched non-Hermitian Luttinger liquids, and interacting Hatano--Nelson systems \cite{DoraMoca2020,TurkeshiSchiro2023,DoraWernerMoca2023,DubeyBiswasKundu2023}. Entanglement dynamics in non-Hermitian systems have further revealed nonequilibrium phase transitions associated with purification and non-Hermitian spectral structures~\cite{PhysRevLett.126.170503,PhysRevX.13.021007}. Collectively, these studies establish non-Hermitian quench dynamics as a distinct many-body problem from that of Hermitian systems.

Parallel developments have brought nonunitary dynamics into contact
with conformal field theory. Recent advances have utilized postselected drives to realize analytically tractable nonunitary CFT evolution \cite{LapierreEtAl2025} and demonstrated that consistent nonunitary conformal boundaries demand independent incoming and outgoing states constructed from dual biorthogonal data \cite{TangWeiWen2026}. Complementary unitary studies have used the integrable field theory generated by a relevant perturbation of the Ising CFT to describe magnetization oscillations and quantum-geometric many-body Landau--Zener dynamics in slowly driven Ising chains~\cite{WangOshikawaKormosWu2024,WangHeWu2025}. Boundary-condition quenches have also been analyzed directly within Lee--Yang field theory \cite{BajnokFulepiLencses2026}.

Despite these advances, a comprehensive boundary conformal field theory (BCFT) framework for global quenches in interacting, bulk non-Hermitian critical systems remains absent. While an ideal framework must microscopically link the left and right initial states to predict local and multipoint observables, a generalized Calabrese-Cardy construction that simultaneously captures complete-character returns, local-primary dynamics, and boundary-sewn spatial correlators has yet to be achieved. Establishing this unified picture is crucial: it would elevate the analysis beyond specific lattice realizations, yielding a universal geometric dictionary that governs nonequilibrium scaling behavior across diverse non-Hermitian quenches.

In this work, we realize this framework by representing the right and left
preparations as the two temporal boundaries of a conformal strip. Assuming both flow to the same conformal boundary condition $B$, their long-distance asymptotic forms are
\begin{align}
 |R_0\rangle
 \propto
 \mathrm{e}^{-\tau_R H_{\rm CFT}}|B\rangle_R,
 \qquad
 \langle L_0|
 \propto
 {}_L\langle B|
 \mathrm{e}^{-\tau_L H_{\rm CFT}},
 \label{eq:intro-paired-boundary}
\end{align}
governed by generally complex one-sided extrapolation parameters $\tau_R$ and $\tau_L$. Linear $\mathcal S$ and antilinear
$\eta$ pairings provide two symmetry-fixed realizations of this
geometry, while the construction itself also allows independent
left and right preparations.  Phase-sensitive projected weights
determine
\begin{equation}
 T=\tau_L+\tau_R,
\end{equation}
while spectroscopy fixes the velocity, and static matrix elements remove any additional vacuum contribution to the
lattice-to-primary operator matching.  These lattice
calibrations are then combined with independently established
BCFT data to determine the dynamical behavior of the observables.

We test this construction in the non-Hermitian Ising chain at the
Yang--Lee critical point, an interacting longitudinal-field spin
model without a free-particle reduction.  The same statically
calibrated boundary data describe the shifted return, the
Yang--Lee-primary one-point function, and the boundary-sewn spatial
correlator.  The two microscopic pairings encode the preparation
phase either as a complex strip modulus or as a displacement on a
real strip.  A direct field-on Yang--Lee quench and a complex
five-state Potts fixed point provide further tests beyond the
filtered benchmark, while independent left and right preparations
are examined in the Appendices.  Together, these results establish a microscopic BCFT framework
for interacting non-Hermitian global quenches across distinct
preparations, observables, and critical theories.

The rest of this paper is structured as follows: Section~\ref{sec:overview} summarizes the static-to-dynamic map and
the principal evidence. Sections~\ref{sec:framework} and \ref{sec:observables} formalize the paired-strip construction, deriving its predictions for return amplitudes alongside one- and two-point correlators. We then apply this machinery to the Yang--Lee and complex Potts models in Secs.~\ref{sec:yl-benchmark} and \ref{sec:potts}. Finally, Sec.~\ref{sec:discussion} outlines the broader physical implications and future extensions of our results. Supporting details, including static calibrations, finite-size checks, and independent left--right preparations, are relegated to the Appendices.

\section{Overview and main results}
\label{sec:overview}

\begin{figure*}[t]
 \includegraphics[width=\textwidth]{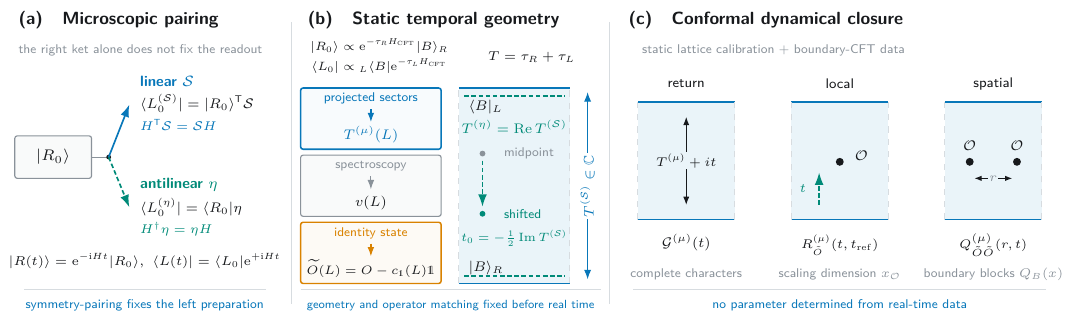}
 \caption{\textbf{Biorthogonal global quenches as temporal-boundary
    problems.}
    (a) A non-Hermitian readout requires both an evolving right ket and a
    left covector.  The linear $\mathcal S$ pairing and the antilinear
    $\eta$ pairing fix two symmetry-related left preparations from the
    same right preparation; the general construction also allows
    independent left and right states.
    (b) In the scaling limit, the preparation pair becomes two temporal
    boundaries with total extrapolation parameter
    $T=\tau_R+\tau_L$.  Projected sector weights determine
    $T^{(\mu)}(L)$, spectroscopy fixes $v(L)$, and an identity-sector
    matrix element fixes the static operator matching $\widetilde O(L)=O-c_{\mathds{1}}(L)\mathds{1}$.
    For a preparation admitting both pairings,
    $T^{(\eta)}=\operatorname{Re}T^{(\mathcal S)}$ and
    $t_0=-\operatorname{Im}T^{(\mathcal S)}/2$, so the same preparation
    phase appears either as a complex strip modulus or as a displacement
    on a real strip.
    (c) These static calibrations, together with independently established
    BCFT data, determine the shifted return, reference-normalized
    one-point function, and one-point-normalized spatial correlator.
    They resolve the complete character content, a primary scaling
    dimension, and the boundary-sewn conformal-block function,
    respectively.}
    \label{fig:concept}
\end{figure*}

Figure~\ref{fig:concept} summarizes the main structure of the formalism.  The
information entering the dynamics separates into static lattice
calibrations and independently established BCFT data:
\begin{align}
 \underbrace{
 \left\{
 T(L),\,
 \tau_R(L),\,
 v(L),\,
 c_{\mathds{1}}(L)
 \right\}
 }_{\text{static lattice calibration}}
 \;+\;
 \underbrace{
 \left\{
 B,\,
 G_i,\,
 x_{\mathcal O},\,
 Q_B
 \right\}
 }_{\text{BCFT data}}\notag\\
 \quad\Longrightarrow\quad
 \left\{
 \widetilde{\mathcal G}(t),\,
 R_{\widetilde{O}}(t,t_{\rm ref}),\,
Q_{\widetilde{O}\widetilde{O}}(r,t)
 \right\}.\notag
\end{align}
Here $B$ is the common infrared conformal boundary condition selected
by the left and right preparations. The coefficients $G_i$ are the
signed or complex coefficients in the character decomposition of the
closed-channel cylinder amplitude associated with $B$;
$x_{\mathcal O}$ is the scaling dimension of the leading nonidentity
primary field $\mathcal O$ appearing in the continuum expansion of the
lattice operator $O$; and $Q_B$ is the one-point-normalized,
boundary-sewn BCFT two-point scaling function associated with $B$.
When the continuum expansion of $O$ contains an identity contribution,
we define the statically matched lattice operator
$\widetilde O(L):=O-c_{\mathds{1}}(L)\mathds{1}$.

The dynamical outputs are defined as follows:
$\widetilde{\mathcal G}$ is the return amplitude with the finite-size
reference energy removed, $R_{\widetilde O}$ is the one-point function
normalized at $t_{\rm ref}$, and
$Q_{\widetilde O\widetilde O}$ is the equal-time spatial correlator
divided by the corresponding one-point functions. Their exact
normalization conventions and strip formulas are given in
Sec.~\ref{sec:observables}. The three outputs probe, respectively, the
complete cylinder character content, the conformal covariance of a
local primary, and the boundary sewing of a spatial two-point function.

\subsection{Temporal geometry from microscopic pairing}

For the symmetry-paired preparations, the linear $\mathcal S$
pairing gives
\begin{equation}
 T^{(\mathcal S)}=2\tau_R,
 \label{eq:overview-S-geometry}
\end{equation}
whereas the antilinear $\eta$ pairing gives
\begin{align}
 T^{(\eta)}
 =&
 2\operatorname{Re}\tau_R=\operatorname{Re}T^{(\mathcal S)},
 \nonumber\\
 t_0
 :=&
 -\operatorname{Im}\tau_R=\operatorname{Im}\tau_L
 =
 -\frac12\operatorname{Im}T^{(\mathcal S)},
 \label{eq:overview-paired-geometries}
\end{align}
where $t_0$ is the part in the one-sided extrapolation parameter that can be absorbed into the time evolution:
\begin{align}
    \vert R(t)\rangle\propto \mathrm{e} ^{-\operatorname{Re}\tau_R  H_{CFT}}\mathrm{e} ^{-i  H_{\rm CFT}(t-t_0)}.
\end{align}
$t_0$ is only properly defined when $\operatorname{Im}\tau_R=-\operatorname{Im}\tau_L$, which is the $\eta$-paired case in this work. The superscript
$\mu\in\{S,\eta\}$ labels the pairing of the preparations.
The same preparation phase is therefore represented either by a
complex strip modulus or by an insertion displacement on a real
strip.

For the Yang--Lee preparations, the independently projected
parameters satisfy
\begin{equation}
 T^{(\eta)}(L)
 =
 \operatorname{Re}T^{(\mathcal S)}(L)
\end{equation}
throughout the sequence $L=12,14,\ldots,22$.  The associated
prediction
\begin{equation}
 t_{0}
 =
 -\frac12\operatorname{Im}T^{(\mathcal S)}
\end{equation}
agrees with the interpolated $\eta$-paired one-point minima for the
product-$Z$ and product-$XZ$ preparations defined in Sec.~\ref{sec:yl-benchmark} within $0.23\%$ and
$0.16\%$, respectively.  The imaginary extrapolation parameter thus
acts as a temporal-boundary datum that can be transferred between
two independently defined microscopic readouts.

\subsection{Conformal closure and scope}

The main Yang--Lee benchmark uses three filtered preparations that
flow to the same identity Cardy boundary but carry different
pairing-resolved temporal parameters.  A static identity subtraction
isolates the leading Yang--Lee primary.  With the boundary condition,
$T$, $v$, and the operator matching fixed, the shifted
return tests the signed character content, the normalized one-point
function tests the dimension $x_\phi=-2/5$ and the insertion
geometry, and the normalized spatial correlator tests the exact
boundary-sewn function
\begin{equation}
 Q_{\widetilde X\widetilde X}(r,t)
 \simeq
 Q_{\mathds{1}}
 \left[
 x(r,t)
 \right].
 \label{eq:overview-block-map}
\end{equation}
Under the $\eta$ pairing, data from all three preparations and the
separations $r=4,6,$ and $8$ collapse, within the observed
finite-size residuals, onto the same real function
$Q_{\mathds{1}}(x)$.  Under the $\mathcal S$ pairing, the temporal
parameters generate different paths in the complex cross-ratio
plane, and the same analytically continued function describes the
corresponding complex correlator trajectories.  The return,
one-point function, and spatial ratio therefore test character
content, local-primary dynamics, and boundary sewing independently.

Three additional comparisons probe the scope of the construction.
In the direct field-on Yang--Lee quench, a real $T$ calibrated from a
Euclidean trajectory predicts held-out Euclidean data and gives
smaller deviations for the local and spatial observables than for
the complete return.  Mixed left and right preparations test the
additivity of independently calibrated one-sided parameters beyond
the symmetry-paired cases.  At the complex five-state Potts fixed
point, a spin-sector calibration predicts both spin and energy
one-point dynamics with different complex dimensions, making the
energy trajectory an independent cross-sector test.

\section{Paired dynamics and complex temporal boundaries}
\label{sec:framework}

In this section, we define the biorthogonal global quench
problem considered in this work, and explain how the Hermitian Calabrese--Cardy ansatz generalizes to
a non-Hermitian strip construction. We first formulate the quench for independent right and left
preparations that flow to the same infrared conformal boundary
condition. In the scaling limit, the two preparations carry
independent, generally complex, one-sided extrapolation parameters.
The resulting strip construction therefore applies beyond any
particular microscopic relation between the two states. For the
main-text lattice tests below, we then specialize to symmetry-paired
preparations, for which the left covector and its one-sided
extrapolation parameter are fixed by the right preparation. This
specialization reduces the nonuniversal input without changing the
underlying strip formulas.

\subsection{Biorthogonal dynamics and paired preparations}

We evolve a right ket and an independently prepared left
covector as
\begin{align}
 \vert{}R(t)\rangle:=e^{-iHt}\vert{}R_0\rangle,
 \qquad
 \langle L(t)\vert{}:=\langle L_0\vert{}e^{+iHt},
 \label{eq:paired-evolution}
\end{align}
where $H$ is the critical non-Hermitian Hamiltonian after the quench.
Equivalently, the ket associated with $\langle L(t)|$ evolves under
$H^\dagger$. The overlap $\langle L(t)|R(t)\rangle$ is conserved.
Provided that it is nonzero, the normalized transition operator
\begin{align}
 \rho_{RL}(t):=
 \frac{\vert{}R(t)\rangle\langle L(t)\vert{}}
      {\langle L(t)\vert{}R(t)\rangle},
 \quad
 i\partial_t\rho_{RL}(t)=[H,\rho_{RL}(t)] .
 \label{eq:rho-rl}
\end{align}
has unit trace and defines normalized expectation values
\begin{align}
 \langle O(t)\rangle:
 =
 \mathrm{Tr}[\rho_{RL}(t)O],
 \label{eq:rl-expectation}
\end{align}
for an operator $O$~\cite{GarrisonWright1988,Brody_2014}.
Since $\rho_{RL}$ need not be Hermitian or positive semidefinite,
these expectation values are generally complex. Hereafter, we omit
the $RL$ label and use $\langle O(t)\rangle$ exclusively in this
biorthogonal sense.

In principle, one can start from any pair of initial states and study their dynamics under the global critical quench. In this work, however, we focus on two specific choices of the left preparation that are directly determined by the right one. If the lattice Hamiltonian admits a nondegenerate symmetric bilinear form $\mathcal{S}$ satisfying
\begin{align}
    H^{\mathsf T} \mathcal{S}
  =
  \mathcal{S} H,
  \qquad
  \mathcal{S}^{\mathsf T}
  =
  \mathcal{S},
\end{align}
we may choose the corresponding linear pairing:
\begin{align}
    \langle L_0\vert{}=\langle L^{(\mathcal{S})}_0\vert{}:=\vert{} R_0\rangle^{\mathsf T}\mathcal{S}.
\end{align}
For the microscopic Hamiltonians used in this work,
$H^{\mathsf T}=H$ in the chosen basis, so that
$\mathcal S=\mathds{1}$ and
\begin{align}
 \langle L_0^{(\mathcal S)}|
 =
 |R_0\rangle^{\mathsf T}.
\end{align}

A second choice is available when the quenched Hamiltonian is
pseudo-Hermitian~\cite{Mostafazadeh2002,Mostafazadeh2004}, with an invertible Hermitian operator $\eta$ satisfying
\begin{align}
 H^\dagger\eta
 &=
 \eta H,
 \qquad
 \eta^\dagger
 =
 \eta .
 \label{eq:eta-pseudo-Hermiticity}
\end{align}
The associated antilinear pairing is
\begin{align}
 \langle L_0^{(\eta)}|
 &:=
 \langle R_0|\eta .
 \label{eq:eta-paired-covector}
\end{align}
The first map is linear in $|R_0\rangle$, whereas the second is
antilinear because $\langle R_0|$ contains complex conjugation.
Since either relation fixes the left preparation once the right one
is specified, we refer to them as symmetry-paired preparations.

\subsection{Complex extrapolation parameters}
To obtain the conformal dynamics on the strip, we generalize the
Calabrese--Cardy ansatz for critical initial states to the
non-Hermitian setting
\cite{CalabreseCardy2006,CalabreseCardy2007,CalabreseCardy2016}.
For a Hermitian critical quench, this ansatz represents a
short-range-entangled initial state by a regularized conformal
boundary state \cite{Ishibashi1989,CardyBoundary1989,Cardy2006BCFT},
\begin{align}
 \vert{}R_0\rangle\propto\mathrm{e}^{-\tau H_{\rm CFT}}\vert{}B\rangle.
 \label{eq:hermitian-preparation}
\end{align}
The Euclidean time $\tau$ is conventionally called the extrapolation
length.

For a non-Hermitian quench with independently specified right and left
preparations, we introduce a separate one-sided extrapolation parameter
on each temporal boundary,
\begin{align}
 \begin{aligned}
  \vert{}R_0\rangle\propto
  \mathrm{e}^{-\tau_RH_{\rm CFT}}\vert{}B\rangle_R,\qquad
  \langle L_0\vert{}\propto
  \langle B\vert{}_L\mathrm{e}^{-\tau_LH_{\rm CFT}},
 \end{aligned}
 \label{eq:general_ansatz}
\end{align}
where $\vert B\rangle_R$ and $\langle B\vert_L$ are right and left boundary states for nonunitary BCFT, and the proportional coefficients denote a normalization fixed by $\langle L_0\vert R_0\rangle=1$~\cite{LiuEtAl2026}. We allow both parameters to be complex with strictly positive real parts ($\operatorname{Re}\tau_R>0$ and $\operatorname{Re}\tau_L>0$), and define their sum,
\begin{align}
T := \tau_L+\tau_R,
\label{eq:total-extrapolation-parameter}
\end{align}
as the total extrapolation parameter of the strip. Following the Hermitian Calabrese--Cardy construction~\cite{Cardy2014,CalabreseCardy2016}, Eq.~\eqref{eq:general_ansatz} represents the leading boundary-RG deformation generated by $H_{\rm CFT}$. We focus on preparations and spacetime regimes for which this
single-$\tau$ description on each temporal boundary captures the
leading infrared behavior, while deformations by additional
irrelevant boundary operators are subleading. Throughout this work, the two preparations are assumed to flow to the
same conformal boundary condition $B$. Independent preparations
flowing to distinct boundary fixed points remain well defined, but
their upper-half-plane representation contains
boundary-condition-changing operators and additional conformal
blocks \cite{CardyBoundary1989,Cardy2006BCFT}. We further restrict to the
finite-size scaling sectors entering the analysis, which admit a
complete biorthogonal eigenbasis, and assume that the corresponding CFT
cylinder Hamiltonian has no Jordan blocks in these sectors.

To separate the Euclidean regularization from the preparation phase, we explicitly decompose each complex one-sided parameter into its Euclidean width and temporal displacement:
\begin{align}
w_R:=\operatorname{Re}\tau_R>0,
 &\qquad
 \delta t_R:=\operatorname{Im}\tau_R.
 \label{eq:one-sided-decomposition}
\end{align}
For example,
\begin{align}
 \mathrm{e}^{-\tau_RH_{\rm CFT}}
 =
 \mathrm{e}^{-w_RH_{\rm CFT}}
 \mathrm{e}^{-\mathrm{i}\delta t_RH_{\rm CFT}},
 \label{eq:complex-filter-decomposition}
\end{align}
and correspondingly for the left preparation.
Thus $w_R$ regularizes the conformal boundary state, while
$\delta t_R$ is a real-time displacement already encoded in the
preparation.

For the symmetry-paired states, the intertwining relations fix the
left one-sided parameter. Using their continuum counterparts,
\begin{align}
 \langle L_0^{(\mathcal S)}|
 &\simeq
 |B\rangle_R^{\mathsf T}
 \mathrm{e}^{-\tau_RH_{\rm CFT}^{\mathsf T}}
 \mathcal S=
 {}_L\langle B|
 \mathrm{e}^{-\tau_RH_{\rm CFT}},
 \label{eq:S-paired-ansatz}
 \\
 \langle L_0^{(\eta)}|
 &\simeq
 |B\rangle_R^\dagger
 \mathrm{e}^{-\tau_R^*H_{\rm CFT}^\dagger}
 \eta
=
 {}_L\langle B|
 \mathrm{e}^{-\tau_R^*H_{\rm CFT}} .
 \label{eq:eta-paired-ansatz}
\end{align}
Here we assume that the pairing maps the right conformal boundary state
to the corresponding left boundary state carrying the same boundary
condition $B$. Eqs.~\eqref{eq:S-paired-ansatz} and
\eqref{eq:eta-paired-ansatz} imply
\begin{align}
 \tau_L^{(\mathcal S)}
 &=
 \tau_R,
 \qquad
 \tau_L^{(\eta)}
 =
 \tau_R^*,
 \label{eq:paired-one-sided-parameters}
\end{align}
and hence
\begin{align}
 T^{(\mathcal S)}
 &=
 2\tau_R,
 \qquad
 T^{(\eta)}
 =
 2\operatorname{Re}\tau_R
 =
 \operatorname{Re}T^{(\mathcal S)}.
 \label{eq:paired-total-parameters}
\end{align}
Hereafter, pairing-dependent quantities carry a superscript
$(\mu)$, where $\mu\in\{\mathcal S,\eta\}$ labels the pairing.
The same preparation phase is therefore represented differently by
the two pairings: it remains in the complex strip modulus for the
$\mathcal S$ pairing, whereas the $\eta$-paired strip has a real total
width and retains $\delta t_R$ as an insertion displacement.

The total parameter can be determined without using real-time data.
Indeed, the Euclidean evolution satisfies
\cite{CardyBoundary1989,Cardy2006BCFT,CalabreseCardy2016}
\begin{align}
 \left\langle
 \mathrm{e}^{-\tau H}
 \right\rangle
 &\propto
 \frac{
 {}_L\langle B|
 \mathrm{e}^{-(T+\tau)H_{\rm CFT}}
 |B\rangle_R
 }{
 {}_L\langle B|
 \mathrm{e}^{-TH_{\rm CFT}}
 |B\rangle_R
 }
 \nonumber\\
 &=
 \frac{Z_B(T+\tau)}{Z_B(T)} .
 \label{eq:euclidean-return-ratio}
\end{align}
Thus projected sector weights or a Euclidean-time trajectory determine
the pairing-dependent total parameter $T^{(\mu)}$ before real-time
evolution, without using any real-time observable. The numerical determinations used below are detailed in
Appendix~\ref{app:T}.

\section{Pairing-resolved conformal observables}
\label{sec:observables}
With the generalized Calabrese-Cardy initial-state ansatz in Eq.~\eqref{eq:general_ansatz}, we are now in a position to derive the universal dynamics of physical observables using boundary conformal field theory (BCFT) on the strip geometry. We begin with the Loschmidt return amplitude, represented by an insertion of real-time evolution, followed by the dynamics of local and spatially separated operators. We restrict our explicit formulas to the $\mathcal S$- and
$\eta$-paired preparations introduced above, for which the left
one-sided parameter is fixed by the right one.

More general independent left and right preparations that flow to the
same Cardy boundary condition obey the same strip formulas, with $\tau_R$ and $T$ treated as independent complex parameters. If the two preparations flow to different boundary fixed points, the strip remains well-defined, but its upper-half-plane representation contains boundary-condition-changing operators and additional conformal blocks. We leave that extension for future work.

\subsection{Return amplitude and the total extrapolation parameter}

The relative-time return amplitude corresponds to inserting an additional real-time evolution operator between the fixed left and right preparations:
\begin{align}
 {\cal G}(t)
 :=\langle e^{-iHt}\rangle.
 \label{eq:general-return}
\end{align}
To evaluate this amplitude within CFT, it is convenient to choose a reference eigenstate $\vert{}\Psi_0\rangle$ and define the shifted return amplitude:
\begin{align}
 \widetilde{\mathcal G}(t)
 &:=
 \left\langle e^{-i[H-E_0(L)]t}\right\rangle
 =
 e^{iE_0(L)t}{\cal G}(t).
\end{align}
With the normalized expectation convention of
Eq.~\eqref{eq:rl-expectation},
$\mathcal G(0)=\widetilde{\mathcal G}(0)=1$.
For a real, energy-ordered spectrum, $\vert{}\Psi_0\rangle$ corresponds to the finite-size ground state. When no natural ground-state ordering exists, we instead choose the state that flows to the CFT identity operator. If the corresponding CFT state has scaling dimension $x_0$, its cylinder energy is
\begin{align}
 E_0^{\rm CFT}(L)
 =
 \frac{2\pi v}{L}
 \left(x_0-\frac{c}{12}\right).
\end{align}
The strip representation then yields
\cite{BloteCardyNightingale1986,Cardy2006BCFT}
\begin{align}
 \widetilde{\mathcal G}(t)
 \simeq
 e^{iE_0^{\rm CFT}(L)t}
 \frac{Z_B(T+it)}{Z_B(T)},
 \qquad
 T=\tau_L+\tau_R,
 \label{eq:general-return-strip}
\end{align}
where $Z_B(T)$ is the unshifted cylinder amplitude with $q(T)=\exp[-4\pi vT/L]$. Note that for the return amplitude, the total extrapolation parameter $T$ is the only relevant parameter; the relative displacement $\tau_R-\tau_L$ between the two temporal boundaries does not enter the expression.

\subsection{One-point functions and the temporal insertion position}
\label{sec:BCFT_one_point}

Unlike the return amplitude, a local operator interrupts the propagation between the two temporal boundaries and therefore resolves how the total extrapolation parameter is divided between the left and right parts.

Let $O$ be a spinless operator of scaling dimension $x_\mathcal{O}$. Its $RL$ expectation value is represented by
\begin{align}\label{eq:one-partition}
 m_O(t)&:=\langle O(t)\rangle\notag\\
 &\propto
 \frac{
 \langle B\vert{}_L
 e^{-[\tau_L-it]H_{\rm CFT}}
 \mathcal{O}
 e^{-[\tau_R+it]H_{\rm CFT}}
 \vert{}B\rangle_R
 }{
 Z_B(T)
 }.
\end{align}
For simplicity, we take the limit $L\to\infty$, where the geometry becomes an infinitely long strip of width $T$. For finite-size systems, this amounts to limiting the prediction to the finite time window $vt\ll L$, where $v$ is the spin velocity. We assume that the microscopic operator $O$ has a single leading
primary component $\mathcal O$ with a nonzero boundary one-point
coefficient for $B$, and that any identity contribution is forbidden
by symmetry or removed by a static operator matching:
\begin{align}
 O(x)
 =
 A_O\,\mathcal{O}(x)+\cdots .
 \label{eq:lattice-primary-matching}
\end{align}
Model-specific operator mixing and finite-lattice corrections are discussed in the corresponding applications.

Setting the right temporal boundary at the origin, the operator is inserted at the complexified Euclidean coordinate
\begin{align}
 u(t)=\tau_R+it
 \label{eq:general-insertion-position}
\end{align}
For a real Euclidean strip, $T>0$ and $0<u<T$, conformal covariance
fixes the one-point function as \cite{McAvityOsborn1995,Cardy2006BCFT,CalabreseCardy2006}
\begin{align}
 m_O(u,T)
 =
 \mathcal{A}_O(T)
 \left[
 \sin\left(
 \frac{\pi u}{T}
 \right)
 \right]^{-x_\mathcal{O}},
 \label{eq:general-onepoint-law}
\end{align}
where $\mathcal{A}_O(T)$ contains the lattice-to-CFT coefficient,
the boundary one-point coefficient, and the overall strip scale.  Eq.~\eqref{eq:general-onepoint-law} is
then analytically continued to complex variables. After removing $\mathcal{A}_O(T)$ by
normalizing at a fixed reference time, we have
\begin{align}
 R_O(t,t_{\rm ref})
 &:=
 \frac{m_O(t)}{m_O(t_{\rm ref})},
 \notag\\
 &\simeq
 \left[
 \frac{
 \sin[\pi(\tau_R+it)/T]
 }{
 \sin[\pi(\tau_R+it_{\rm ref})/T]
 }
 \right]^{-x_\mathcal{O}}.
 \label{eq:general-onepoint-ratio}
\end{align}
For complex $\tau_R$, $T$, or $x_{\mathcal O}$, the power on the right-hand side is defined through a logarithm
continued continuously along the time path and anchored at
$t=t_{\rm ref}$, where $R_O(t_{\rm ref},t_{\rm ref})=1$.

The above general scaling of the local observables can be simplified by considering the paired preparations. For the linear $\mathcal S$-bilinear pairing, the insertion sits at the midpoint of the strip, $\tau_R=T/2$, so Eq.~\eqref{eq:general-onepoint-ratio} reduces to
\begin{align}
    R_O^{(\mathcal S)}(t,t_{\rm ref})
 \simeq
 \left[
 \frac{
 \cosh(\pi t/T)
 }{
 \cosh(\pi t_{\rm ref}/T)
 }
 \right]^{-x_\mathcal{O}}.
 \label{eq:S-onepoint-ratio}
\end{align}
Because $T$ may be complex, even a real scaling dimension can generate a nontrivial phase trajectory. When both $T$ and $x_\mathcal{O}$ are complex, their real and imaginary parts generally contribute to both the magnitude and phase.

For the antilinear $\eta$-pseudo-Hermitian pairing, although the insertion is not at the midpoint, the imaginary part $\delta t_R$ can be absorbed into the real-time evolution. Thus, Eq.~\eqref{eq:general-onepoint-ratio} reduces to
\begin{align}
R_O^{(\eta)}(t,t_{\rm ref})
 \simeq
 \left[
 \frac{
 \cosh[\pi(t+\delta t_R)/T]
 }{
 \cosh[\pi(t_{\rm ref}+\delta t_R)/T]
 }
 \right]^{-x_\mathcal{O}}.
 \label{eq:eta-onepoint-ratio}
\end{align}
Therefore, the preparation phase does not produce a complex strip modulus in the $\eta$-paired geometry. Instead, it shifts the temporal center of the one-point expectation value to
\begin{align}
 t_0\equiv-\delta t_R.
 \label{eq:eta-temporal-center}
\end{align}
While this shift can be represented mathematically by the replacement $t\mapsto t+\delta t_R$, it acquires physical meaning once the microscopic quench fixes the time origin. If the same right preparation also admits the linear $\mathcal S$ pairing, then
\begin{align}\label{eq:delayvsima}
 t_0
 =
 -\frac{\operatorname{Im}T^{(\mathcal S)}}{2},
\end{align}
so the temporal center of the $\eta$-paired observable is predicted by the imaginary part of the $\mathcal S$-paired extrapolation parameter. Because $T^{(\mathcal S)}$ and $t_0$ are obtained in distinct
microscopic pairing protocols, Eq.~\eqref{eq:delayvsima} provides a
direct dynamical test of the paired-boundary relation in
Eq.~\eqref{eq:paired-total-parameters}.

For real $T$, $\delta t_R$, and scaling dimension $x_\mathcal{O}$, the normalized ratio in
Eq.~\eqref{eq:eta-onepoint-ratio} is real. A primary with $x_\mathcal{O}>0$ has a maximum at $t=t_0$ and decays exponentially away from this center at late times. A negative scaling dimension reverses this behavior: the profile has a minimum at the shifted center and grows as $\vert{}t-t_0\vert{}$ increases. The $\eta$ pairing therefore retains the usual real-strip scaling form, but the microscopic preparation determines where that form is centered in real time.
\subsection{Two-point functions and boundary conformal blocks}
\label{sec:general-two-point}

A two-point function contains information beyond the scaling dimension of a single operator. After the one-point responses of the two insertions are removed, the remaining dependence is a universal function of the strip cross ratio. This function is fixed by the allowed intermediate conformal families and by the boundary sewing data.

We consider the same strip geometry as in Eq.~\eqref{eq:one-partition}, with two operators spatially separated by $r$:
\begin{align}
    G_{OO}(r,t):=\langle O_iO_{i+r}\rangle.
\end{align}
Their strip coordinates are
\begin{align}
 \zeta_{1,2}
 =
 \mp\frac{r}{2v}
 +iu(t),
 \label{eq:strip-two-point-coordinates}
\end{align}
where $v$ is the spin velocity. For a real Euclidean strip, where $T>0$ and $0<\tau_R<T$, conformal invariance fixes the correlator to
\cite{McAvityOsborn1995,Cardy2006BCFT,Runkel1999}
\begin{align}\label{eq:2-point-mmF}
    G_{OO}(r,u,T)\simeq m^2_O(u,T)Q_B[x(r,u,T)],
\end{align}
where $x$ is the cross ratio:
\begin{align}
    x(r,u,T):=\frac{\cosh [\pi r/(vT)]-1}{\cosh[\pi r/(vT)]-\cos(2\pi u/T)}
\end{align}
and $Q_B(x)$ is the one-point-normalized BCFT scaling function. After
the kinematic factors are removed, it depends only on the cross ratio
and admits the bulk-channel expansion
\begin{align}
    Q_B(x)\propto(1-x)^{x_\mathcal{O}}\sum_{p\in\mathcal{O}\times\mathcal{O}}\lambda_p^{(B)}\mathcal{F}_p(x),
    \label{eq:Q_B}
\end{align}
where the sum runs over the conformal blocks $\mathcal F_p(x)$ with sewing coefficients $\lambda_p^{(B)}$. The overall proportionality factor in Eq.~\eqref{eq:Q_B} is fixed by the
one-point normalization in Eq.~\eqref{eq:2-point-mmF}. When the identity is the unique
leading contribution in the relevant boundary channel, this definition
implies
\begin{align}
 \lim_{x\to1} Q_B(x)=1.
\end{align}
Analytically continuing Eq.~\eqref{eq:2-point-mmF} to the complex parameters now gives
\begin{align}\label{eq:G_OO=Q_B}
    \frac{G_{OO}(r,t)}{m^2_O(t)}\simeq Q_B[x(r,t)],
\end{align}
where
\begin{align}\label{eq:cross-ratio}
    x(r,t):=\frac{\cosh [\pi r/(vT)]-1}{\cosh[\pi r/(vT)]-\cos(2\pi (\tau_R+it)/T)}.
\end{align}
One may also consider the connected part:
\begin{align}\label{eq:connected-correlator}
    C_{OO}(r,t):=\langle O_{0}O_r\rangle-\langle O\rangle^2\simeq m^2_O(t)(Q_B[x(r,t)]-1).
\end{align}

For the $\mathcal{S}$ pairing, the insertions remain at the midpoint of the complexified strip, and Eq.~\eqref{eq:cross-ratio} reduces to
\begin{align}\label{eq:cross-ratio-S}
 x_\mathcal{S}(r,t)
 =
 \frac{
 \cosh[\pi r/(vT^{(\mathcal{S})})]-1
 }{
 \cosh[\pi r/(vT^{(\mathcal{S})})]
 +
 \cosh(2\pi t/T^{(\mathcal{S})})
 }.
\end{align}
For a complex extrapolation parameter, $x_\mathcal{S}$ follows a preparation-dependent path in the complex plane. The Euclidean branch of the same boundary-sewn function $Q_B$ must
then be analytically continued continuously along this path.

Thus, the preparation phase affects the spatial correlator in two ways: through the complex one-point factor $m_O(t)^2$ and through the relative phase accumulated between conformal blocks along the complex cross-ratio path.

For the antilinear $\eta$ pairing, the imaginary part of the extrapolation parameter once again enters as a shift of the time origin:
\begin{align}
 x^{(\eta)}(r,t)
 =
 \frac{
 \cosh[\pi r/(vT^{(\eta)})]-1
 }{
 \cosh[\pi r/(vT^{(\eta)})]
 +
 \cosh[2\pi(t+\delta t_R)/T^{(\eta)}]
 }.
 \label{eq:eta-cross-ratio}
\end{align}
For real $r$, $t$, $v$, and $T^{(\eta)}$, this cross ratio remains strictly within the interval $0<x^{(\eta)}<1$. The preparation phase therefore does not generate a complex conformal-block path.

\section{Conformal dynamics in the non-Hermitian Ising chain}
\label{sec:yl-benchmark}
The non-Hermitian Ising chain provides a particularly useful benchmark because the same microscopic Hamiltonian realizes both pairing structures introduced in Sec.~\ref{sec:framework}. This allows us to compare the two temporal geometries without altering the post-quench Hamiltonian or the prepared right state. The $\mathcal{S}$ pairing represents the preparation via a midpoint insertion on a generally complex strip, whereas the $\eta$ pairing represents the same preparation via an off-center insertion on a real strip. We first determine these geometries using static projected data and then hold them fixed in all real-time comparisons.

\subsection{Lattice Hamiltonian and filtered preparations}

We study conformal quench dynamics in the non-Hermitian Ising chain
\begin{align}
 H_{\rm YL}=-\sum_{j=1}^{L}
 \left[Z_jZ_{j+1}+\lambda X_j+\mathrm{i} h_zZ_j\right],
 \qquad Z_{L+1}=Z_1 .
 \label{eq:yl-hamiltonian}
\end{align}
The Hamiltonian is invariant under the antiunitary symmetry
$P\mathcal K$, where
\begin{align}
 P:=\prod_j X_j
\end{align}
is the parity operator and $\mathcal K$ denotes complex conjugation. For each
$\lambda>1$, tuning the imaginary longitudinal field to the
thermodynamic edge $h_z=h_z^c(\lambda)$ realizes the nonunitary
$M(2,5)$ Yang--Lee CFT
\cite{YangLee1952,LeeYang1952,Fisher1978,Cardy1985,vonGehlen1991}.
In the computational basis,
\begin{align}
 H_{\rm YL}^{\mathsf T}
 &=
 H_{\rm YL},
 \qquad
 H_{\rm YL}^{\dagger}P
 =
 PH_{\rm YL}.
 \label{eq:yl-two-pairings}
\end{align}
The first relation realizes the linear pairing with
$\mathcal S=\mathds{1}$, while the second realizes the
pseudo-Hermitian pairing with $\eta=P$.

Throughout the Yang--Lee calculations, we set
$\lambda=4$ and
$h_z=h_z^c(4)\simeq1.55872195$ as determined in
Ref.~\cite{LiuEtAl2026}. At this thermodynamic critical field, the
finite-size low-energy sectors used below remain in the unbroken
$P\mathcal K$ regime. Their eigenvalues, and in particular the gap
between the $\phi$ and identity representatives, are therefore real.

The initial states used here are referred to as filtered product states. Starting from three translationally invariant product states:
\begin{align}
&\vert{}X\rangle:=\left(\frac{\vert{}\uparrow\rangle+\vert{}\downarrow\rangle}{\sqrt{2}} \right)^{\otimes L},\quad\vert{}Z\rangle:=\vert{}\uparrow\rangle^{\otimes L},\quad\notag
 \\
 &\vert{}XZ\rangle:=\left[\cos\!\left(\frac{\pi}{8}\right)\vert{}\uparrow\rangle
 +\sin\!\left(\frac{\pi}{8}\right)\vert{}\downarrow\rangle\right]^{\otimes L},
 \label{eq:yl-real-preparations}
\end{align}
we apply a short Euclidean filter to suppress ultraviolet lattice corrections by defining
\begin{align}
 \begin{aligned}
 \vert{}R_\alpha\rangle
 &:=\mathrm{e}^{-\frac{\beta}2
 [H_{\rm YL}-E_0(L)]}\vert{}\alpha\rangle,\quad \alpha\in\{X,Z,XZ\},
 \end{aligned}
 \label{eq:yl-filtered-preparation}
\end{align}
where $E_0(L)$ is the ground-state energy~\cite{CarignanoTagliacozzo2025,RobertsonSuraceTagliacozzo2022}. The filtering strength is set to $\beta=1$ throughout this section; thus, we omit the explicit dependence on $\beta$ and refer to $|R_\alpha\rangle$ as the product-$\alpha$ preparation. The corresponding paired covectors are
\begin{align}
 \langle L_\alpha^{(\mathcal S)}\vert{}
 :=
 \vert{}R_\alpha\rangle^{\mathsf T},
 \qquad
 \langle L_\alpha^{(\eta)}\vert{}
 :=
 \langle R_\alpha\vert{}P .
 \label{eq:yl-paired-covectors}
\end{align}
The bare product-$X$ state is real and parity even. Since the
Euclidean filter preserves the antiunitary $P\mathcal K$
symmetry,
\begin{align}
 P\mathcal K|R_X\rangle=|R_X\rangle .
\end{align}
Consequently,
\begin{align}
 |R_X\rangle^{\mathsf T}
 =
 \langle R_X|P,
\end{align}
and the two paired covectors coincide. The filtered product-$Z$
and $XZ$ preparations do not obey this invariance and therefore
give distinct left covectors under the two pairings.

Periodic-chain spectroscopy identifies all these preparation pairs with the identity Cardy boundary. In the main text, we focus on the results for the paired product preparations introduced above. Additional left preparations interpolating between the two symmetry-paired covectors are discussed in Appendix~\ref{app:mixed-preparations}.
\subsection{Pairing-resolved extrapolation parameters}
\label{sec:yl-extrapolation-parameters}
For general preparations $\vert{}R_0\rangle$ and $\langle L_0\vert{}$ that flow to the same boundary condition, we define the ratio
\begin{align}
 {\mathcal{R}}(L)
 :=-\frac{1}{\varphi}
 \frac{\langle L_0\vert{}
 \Psi_1^R\rangle\langle\Psi_1^L
 \vert{}R_0\rangle
 }{
 \langle L_0\vert{}
 \Psi_0^R\rangle\langle\Psi_0^L
 \vert{}R_0\rangle
 } ,
 \label{eq:yl-projected-ratio}
\end{align}
where $|\Psi_n^R\rangle$ and $|\Psi_n^L\rangle$ are the $n$th right and left eigenvectors of $H_{\rm YL}$ satisfying $\langle\Psi_n^L\vert{}\Psi_{n^\prime}^R\rangle=\delta_{n n^\prime}$. In the CFT limit, $\Psi_0$ and $\Psi_1$ flow to $\phi$ and identity, respectively. The coefficient $-\frac{1}{\varphi}$ follows from independently known universal ratio of boundary coefficients $G_{\mathds{1}}/G_\phi=-\varphi$ and $\varphi=(1+\sqrt5)/2$. Following the general discussion in Appendix~\ref{app:T}, the finite-size total parameter $T(L)$ can be determined as
\begin{align}
 T(L)
 =
 -\frac{
 \operatorname{Log}_{\rm cont}{\mathcal{R}}(L)
 }{
 E_1(L)-E_0(L)
 }.
 \label{eq:yl-projected-T}
\end{align}

For the real and nondegenerate Yang--Lee sectors considered here, the combined complex-symmetric and pseudo-Hermitian structures imply
\begin{align}
 {\mathcal{R}}^{(\eta)}(L)
 =
 \left\vert{}
 {\mathcal{R}}^{(\mathcal S)}(L)
 \right\vert{}
\end{align}
for any right preparation together with either symmetry-paired left
covector. The real finite-size gap then implies
\begin{align}
 T_{\alpha}^{(\eta)}(L)
 =
 \operatorname{Re}
 T_{\alpha}^{(\mathcal S)}(L).
 \label{eq:yl-T-relation}
\end{align}

\begin{figure}[t]
 \includegraphics[width=\columnwidth]{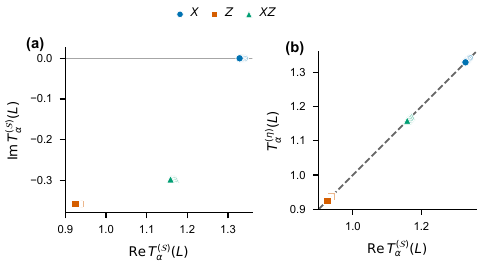}
 \caption{
 \textbf{Pairing-resolved total extrapolation parameters in the
 Yang--Lee chain.}
 (a) Complex-plane trajectories of $T_{\alpha}^{(\mathcal S)}$ for the three filtered product preparations at $\beta=1$ and $L=12,14,\ldots,22$. Product $X$ remains real, whereas product $Z$ and $XZ$ preparations retain finite imaginary components. Symbols darken with increasing $L$.
(b) Independently extracted
$T_\alpha^{(\eta)}(L)$ versus
$\operatorname{Re}T_\alpha^{(\mathcal S)}(L)$.
Dashed diagonal represents
the analytic identity in Eq.~\eqref{eq:yl-T-relation}.}
\label{fig:yl-pairing-parameters}
\end{figure}
Figure~\ref{fig:yl-pairing-parameters}(a) shows the finite-size flow
of the three $\mathcal S$-paired parameters. Product $X$ remains on
the real axis, whereas the product-$Z$ and product-$XZ$ preparations retain
stable imaginary components. Because Eq.~\eqref{eq:yl-T-relation} follows analytically from the
real, nondegenerate sector structure, Fig.~\ref{fig:yl-pairing-parameters}(b) is an independent
numerical consistency check rather than an additional fit or
prediction. The $\mathcal S$- and $\eta$-paired projected ratios are
processed separately, yet the resulting parameters satisfy the
diagonal relation to numerical precision throughout the finite-size
sequence.

The finite-size spin velocity is defined via the finite-size energy gap
\cite{BloteCardyNightingale1986}:
\begin{align}\label{eq:yl-velocity}
    v(L):=\frac{L\left[E_1(L)-E_0(L)\right]}{2\pi (0-x_\phi)}.
\end{align}
Eq.~\eqref{eq:yl-projected-T} operationally defines a finite-size effective total extrapolation parameter. At finite $L$, the projected lattice weights contain corrections from irrelevant bulk and boundary operators, so $T_{\alpha}^{(\mu)}(L)$ may not yet equal its thermodynamic-limit value~\cite{LiuEtAl2026}.

Accordingly, each finite-size dynamical trajectory is compared with
a prediction constructed from
$T_{\alpha}^{(\mu)}(L)$ and $v(L)$ determined independently at
that same system size. Exploiting translation invariance, we reach the largest accessible
size $L=22$. The total parameters extracted at this size, including
the $\beta=1$ filter, are
\begin{align}
\begin{array}{c|cc}  \alpha & T_{\alpha}^{(\mathcal S)} & T_{\alpha}^{(\eta)} \\ \hline  X & 1.32865 & 1.32865 \\  Z & 0.92510-0.35807\,i & 0.92510 \\  XZ & 1.15885-0.29744\,i & 1.15885 \end{array}\notag
\end{align}
All the following Yang--Lee dynamical comparisons are performed at
this system size. For notational economy, we henceforth write
\begin{align}
 T_{\alpha}^{(\mu)}
 &\equiv T_{\alpha}^{(\mu)}(22),
 \qquad
 v\equiv v(22),
\end{align}
unless the system-size dependence is displayed explicitly. The
finite-size flow of these quantities and the effect of replacing
$T_{\alpha}^{(\mu)}(22)$ by an asymptotic estimate are analyzed in
Appendix~\ref{app:finite-size-T}.

\subsection{Return amplitudes and local
operator dynamics}
\label{sec:yl-global-local}

The Cardy boundary condition and the pairing-resolved extrapolation
parameters have now been fixed independently of real-time evolution.  Before
testing local dynamics, we also match the lattice operator to the Yang--Lee
primary.  The bare transverse-field operator $X$ contains an allowed identity
component, which we determine from the finite-size state flowing to the CFT
identity~\cite{ZouMilstedVidal2020}:
\begin{align}
 c_{\mathds{1}}(L)
 &:={}
 {
 \langle\Psi_{\mathds{1}}^L|X_j|\Psi_{\mathds{1}}^R\rangle
 },
 \nonumber\\
 \widetilde X_j(L)
 &:={}
 X_j-c_{\mathds{1}}(L)\mathds{1},
 \label{eq:yl-identity-subtracted-X}
\end{align}
where
$|\Psi_{\mathds{1}}^R\rangle$ and
$\langle\Psi_{\mathds{1}}^L|$
are the finite-size right and left representatives flowing to the
CFT identity.
Translation invariance makes $c_{\mathds{1}}(L)$ independent of $j$.  It is
fixed by the final Hamiltonian and is independent of the preparation, pairing and real-time data.  At $L=22$,
\begin{align}
 c_{\mathds{1}}(22)\approx0.05750468,
 \qquad
 |\operatorname{Im}c_{\mathds{1}}(22)|<7\times10^{-11},
\end{align}
where the full complex value is retained in the calculation.  We henceforth
write $\widetilde X_j\equiv\widetilde X_j(22)$. After this subtraction, the leading scaling component of $\widetilde X$ is the Yang--Lee primary $\phi$ with
$x_\phi=-2/5$.  A detailed discussion of this subtraction can be found in Appendix~\ref{app:yl-operator-matching}.

The shifted lattice return amplitude follows directly from Eq.~\eqref{eq:general-return-strip}:
\begin{align}
 {\widetilde{\mathcal G}(t)}
\simeq
e^{-\frac{i\pi vt}{15L}}
\frac{Z_\mathds{1}[T+it]}
     {Z_\mathds{1}[T]},
 \label{eq:yl-paired-return}
\end{align}
where
\begin{align}
 Z_\mathds{1}(T)\propto
\chi_\phi[q(T)]
-
\varphi\chi_{\mathds{1}}[q(T)]
 \label{eq:yl-return-character}
\end{align}
is the cylinder partition function with identity Cardy boundary on both sides
\cite{CardyBoundary1989,DoreyEtAl2000,LiuEtAl2026}.

For product-$X$ preparation, coincidence of the two paired covectors implies
\begin{align}
 \widetilde{\mathcal G}_{X}^{(\mathcal S)}(t)
 =
 \widetilde{\mathcal G}_{X}^{(\eta)}(t)
 \label{eq:yl-X-return-identity}
\end{align}
explicitly on the lattice. This equality also holds for any expectation value under product-$X$ preparation.

\begin{figure*}[t]
 \includegraphics[width=\textwidth]{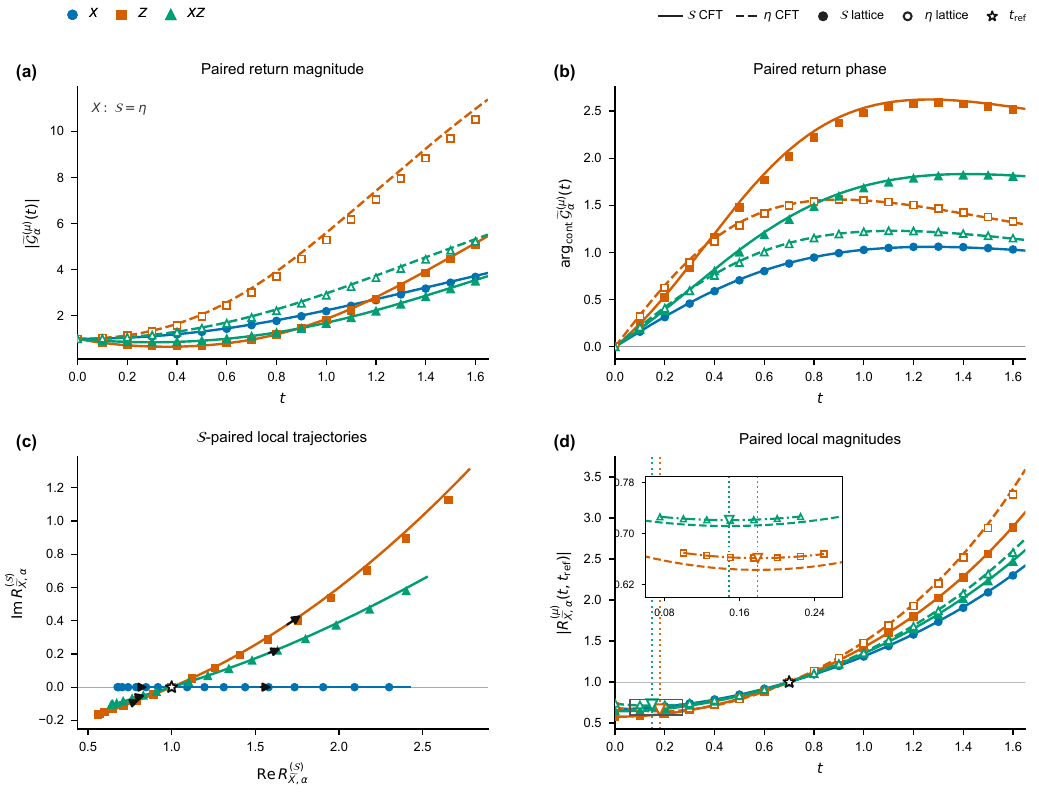}
 \caption{
\textbf{Return amplitudes and local-primary dynamics for paired preparations in the Yang--Lee model.}
(a,b) Magnitude and continuously tracked phase of the return amplitudes at
$L=22$ for product $X$, product $Z$, and the $XZ$ preparation.  Solid curves
and filled symbols denote the $\mathcal S$ pairing; dashed curves and open
symbols denote the $\eta$ pairing.  Curves are the signed complete-character
predictions using the statically determined extrapolation parameters, and
symbols are lattice data.  The two product-$X$ returns coincide exactly.
(c) Complex trajectories of the normalized $\mathcal S$-paired one-point
 function of $\widetilde X_j$.  Arrows indicate increasing time, and
 all trajectories pass through $R_{\widetilde X}(t_{\rm ref})=1$.
(d) Magnitudes of the normalized $\mathcal S$- and $\eta$-paired one-point
functions.  The vertical dotted lines mark the static predictions
 {$t_{\rm min}=t_0
=-\operatorname{Im}T_\alpha^{(\mathcal S)}/2$};
triangular markers show the lattice minima obtained by seven-point local quartic
interpolation.  The enlarged inset shows the interpolants as short
dash-dotted segments over the boxed early-time region.}
\label{fig:yl-global-local}
\end{figure*}

Figures~\ref{fig:yl-global-local}(a) and~\ref{fig:yl-global-local}(b) display the magnitude and continuously tracked phase of both paired returns at $L=22$. Product $X$ lies on the common trajectory of the two pairings. The product-$Z$ and product-$XZ$ preparations exhibit distinct pairing-dependent magnitude and phase evolution, both accurately described by Eq.~\eqref{eq:yl-paired-return} using their statically determined parameters.

We next turn to the local-primary dynamics.  Define
\begin{align}
 m_{\widetilde X,\alpha}^{(\mu)}(t)
 :=
 \langle\widetilde X_j(t)\rangle_{\alpha}^{(\mu)},
 \qquad
 R_{\widetilde X,\alpha}^{(\mu)}(t,t_{\rm ref})
 :=
 \frac{
 m_{\widetilde X,\alpha}^{(\mu)}(t)
 }{
 m_{\widetilde X,\alpha}^{(\mu)}(t_{\rm ref})
 },
 \label{eq:yl-subtracted-onepoint}
\end{align}
where the same fixed reference time $t_{\rm ref}=0.7$ is used for all preparations and pairings.

For the linear $\mathcal S$ pairing, the operator is inserted at the midpoint of the generally complex strip. Eq.~\eqref{eq:S-onepoint-ratio} yields
\begin{align}
 R_{\widetilde X,\alpha}^{(\mathcal S)}
 (t,t_{\rm ref})
 \simeq
 \left[
 \frac{
 \cosh[\pi t/T_{\alpha}^{(\mathcal S)}]
 }{
 \cosh[\pi t_{\rm ref}/
 T_{\alpha}^{(\mathcal S)}]
 }
 \right]^{-x_\phi}.
 \label{eq:yl-S-onepoint}
\end{align}
The negative scaling dimension drives the growth of this expectation value. For the product-$Z$ and product-$XZ$ preparations, where $T_{\alpha}^{(\mathcal S)}$ is complex, the same formula also captures the observed phase evolution.

For the antilinear $\eta$ pairing, the corresponding real-strip prediction is
\begin{align}
 R_{\widetilde X,\alpha}^{(\eta)}
 (t,t_{\rm ref})
 \simeq
 \left[
 \frac{
 \cosh\!\left[
 \pi(t-t_{0,\alpha})/
 T_{\alpha}^{(\eta)}
 \right]
 }{
 \cosh\!\left[
 \pi(t_{\rm ref}-{t_{0,\alpha}})/
 T_{\alpha}^{(\eta)}
 \right]
 }
 \right]^{-x_\phi},
 \label{eq:yl-eta-onepoint}
\end{align}
where $t_{0,\alpha}$ is the effective time shift introduced by the complexity of the extrapolation parameter $\tau_{R}$.
For $x_\phi=-2/5$, Eq.~\eqref{eq:yl-eta-onepoint} has a minimum at
\begin{align}
 {t_{\rm min}}
 =t_{0,\alpha}=-\frac{
 \operatorname{Im}T_{\alpha}^{(\mathcal S)}}
 {2},
 \label{eq:yl-eta-minimum}
\end{align}
where the last equality follows from Eq.~\eqref{eq:paired-total-parameters}. As discussed at the end of Sec.~\ref{sec:BCFT_one_point}, the static parameter $T^{(\mathcal{S})}$ therefore predicts the location of the minimum of $\langle\widetilde{X}\rangle$. This then serves as a non-trivial check of the relation between paired extrapolation parameters~\eqref{eq:paired-one-sided-parameters}.

Figure~\ref{fig:yl-global-local}(c) displays the $\mathcal S$-paired
one-point ratios as trajectories in the complex plane. Product $X$ remains
on the real axis, while the product-$Z$ and product-$XZ$ preparations follow distinct
complex trajectories fixed by their respective extrapolation parameters.
Fig.~\ref{fig:yl-global-local}(d) compares the magnitudes of the
$\mathcal S$- and $\eta$-paired ratios with the CFT predictions. For the product-$Z$ and product-$XZ$ preparations, the
vertical lines mark the predicted centers in Eq.~\eqref{eq:yl-eta-minimum},
without fitting a temporal displacement.  The statically predicted and dynamical lattice
minima are $0.179035$ and $0.179449$ for product $Z$, and $0.148718$ and
$0.148962$ for the $XZ$ preparation, corresponding to relative differences
of $0.23\%$ and $0.16\%$, respectively.

The return amplitude and the one-point function therefore probe complementary aspects of the paired temporal geometry. The return amplitude tests the signed complete-character continuation at the pairing-specific total parameter, whereas the local operator resolves how the same preparation phase is represented as a complex strip modulus under the $\mathcal S$ pairing and as a temporal displacement under the $\eta$ pairing.

 \begin{figure*}[t]
 \includegraphics[width=\textwidth]{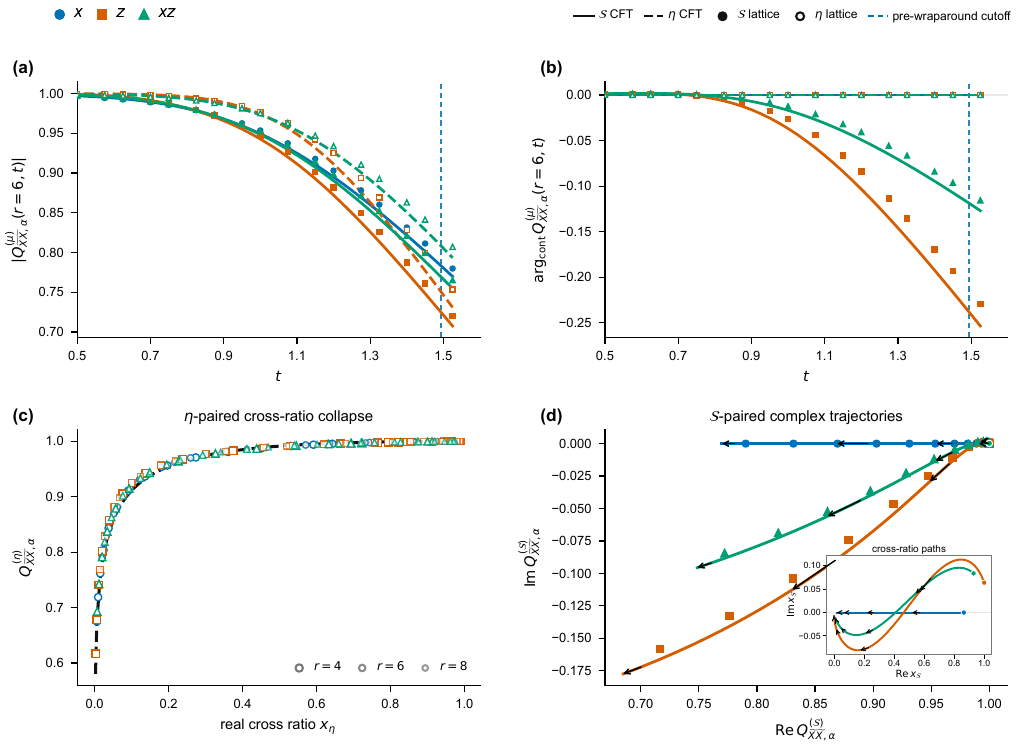}
 \caption{\textbf{Correlator dynamics for filtered paired preparations in the Yang--Lee model.}
All lattice data use $L=22$ and the statically subtracted operator in
Eq.~\eqref{eq:yl-identity-subtracted-X}.  The plotted ratio is
$Q_{\widetilde X\widetilde X,\alpha}^{(\mu)}(r,t)=
G_{\widetilde X\widetilde X,\alpha}^{(\mu)}(r,t)/
[m_{\widetilde X,\alpha}^{(\mu)}(t)]^2$, without an additional time
normalization. Colors and marker shapes identify the preparations $X$, $Z$,
and $XZ$; solid curves and filled symbols denote the $\mathcal S$ pairing,
while dashed curves and open symbols denote the $\eta$ pairing.
 (a) $\vert{}Q_{\widetilde X\widetilde X}\vert{}$ and (b) its continuously unwrapped phase at $r=6$.
 The vertical dashed line marks the conservative
 pre-wraparound condition in Eq.~\eqref{eq:yl-prewraparound}. Early-time data ($t<0.5$) remain nearly constant and are therefore omitted.
 (c) $\eta$-paired data for $r=4,6,$ and $8$ collapse onto the real
 cross-ratio function of $x_{\alpha}^{(\eta)}$; the local legend identifies the
 three separations.
 (d) $\mathcal S$-paired complex trajectories at $r=6$, with black arrows
 indicating increasing time and the inset showing the corresponding
 complex cross-ratio paths.}
 \label{fig:yl-bpz}
 \end{figure*}

 \subsection{Spatial correlations and boundary conformal blocks}

Finally, we test the conformal dynamics of the spatial two-point function. The general universal behavior was derived in Sec.~\ref{sec:general-two-point}. To fully determine the spatial correlator, one must specify the boundary-sewn two-point function $Q_B$, which requires almost complete knowledge of the field theory~\cite{lewellenSewingConstraintsConformal1992}. For the Yang--Lee CFT, due to its minimal structure, this function can be determined explicitly~\cite{Runkel1999}.

The Yang--Lee primary $\phi$, with $h_\phi=-1/5$, obeys the
level-two null-state condition
\begin{align}
 \left[L_{-2}-\frac52L_{-1}^2\right]\vert{}\phi\rangle=0.
 \label{eq:yl-null-state}
\end{align}
For a chiral block $f(x)$ of
$\langle\phi(\infty)\phi(1)\phi(x)\phi(0)\rangle$, the null-state
condition yields
\begin{align}
 f''(x)+\frac{2(2x-1)}{5x(x-1)}f'(x)
 +\frac{2}{25x^2(x-1)^2}f(x)=0.
 \label{eq:yl-bpz-ode}
\end{align}
A convenient basis of bulk-channel Virasoro blocks is
\cite{BelavinPolyakovZamolodchikov1984,DotsenkoFateev1984,Cardy1985}
\begin{align}
 \mathcal F_{\mathds{1}}(x)
 &=[x(1-x)]^{2/5}\,
 {}_2F_1\left(\frac35,\frac45;\frac65;x\right),
 \nonumber\\
 \mathcal F_{\phi}(x)
 &=x^{1/5}(1-x)^{2/5}\,
 {}_2F_1\left(\frac25,\frac35;\frac45;x\right).
 \label{eq:yl-bpz-blocks}
\end{align}
Their leading powers as $x\to0$ resolve the two bulk intermediate channels in the fusion rule
\begin{align}\label{eq:fusion}
 \phi\times\phi=\mathds{1}+\phi.
\end{align}

For the identity Cardy boundary, the boundary-preserving
open-channel spectrum is
$\mathcal H_{\mathds{1}\mathds{1}}=\mathcal V_{\mathds{1}}$.
Hence the $x\to1$ boundary-channel expansion contains only the
identity Virasoro family. Crossing expresses this single boundary
block as a fixed combination of the two bulk-channel blocks allowed
by the fusion rule~\eqref{eq:fusion}
\cite{CardyBoundary1989,Runkel1999,DoreyEtAl2000}:
\begin{align}
 \begin{aligned}
 K_1(x)
 &=
 -\mathcal F_{\mathds{1}}(x)
 +\frac{A_{\rm YL}^2}{\varphi}\mathcal F_\phi(x)
 \\
 &=
 \frac{1}{\varphi}[x(1-x)]^{2/5}
 {}_2F_1\left(\frac35,\frac45;\frac65;1-x\right),
 \\
 A_{\rm YL}^2
 &=
 \frac{\Gamma(1/5)\Gamma(6/5)}
      {\Gamma(3/5)\Gamma(4/5)},
 \qquad
 \varphi=\frac{1+\sqrt5}{2}.
 \end{aligned}
 \label{eq:yl-physical-block}
\end{align}
The first line resolves the identity and $\phi$ bulk channels, whereas the second line represents the pure identity block in the boundary-channel expansion. Since the squared bulk-to-boundary identity coefficient is
\begin{align}
 \left(B_1^{\phi\mathds{1}}\right)^2=\frac1\varphi,
\end{align}
the normalized function introduced in Sec.~\ref{sec:general-two-point} is
\begin{align}
 Q_\mathds{1}(x)
 &=
 \frac{K_1(x)}
 {\left(B_1^{\phi\mathds{1}}\right)^2(1-x)^{2/5}}
 \nonumber\\
 &=
 x^{2/5}\,
 {}_2F_1\left(\frac35,\frac45;\frac65;1-x\right),
 \qquad
 \lim_{x\to1}Q_\mathds{1}(x)=1.
 \label{eq:yl-Q1}
\end{align}
Thus, boundary sewing fixes the relative weight of the two bulk blocks, while division by the squared one-point coefficient ensures the correct normalization at $x=1$.

We now apply Eq.~\eqref{eq:yl-Q1} to the lattice operator
$\widetilde X_j$. For $\mu=\mathcal{S},\eta$, define
\begin{align}
 G_{\widetilde X\widetilde X,\alpha}^{(\mu)}(r,t)
 &:=
 \left\langle \widetilde X_i(t)\widetilde X_{i+r}(t)
 \right\rangle_{\alpha}^{(\mu)},
 \nonumber\\
 Q_{\widetilde X\widetilde X,\alpha}^{(\mu)}(r,t)
 &:=
 \frac{G_{\widetilde X\widetilde X,\alpha}^{(\mu)}(r,t)}
 {\left[m_{\widetilde X,\alpha}^{(\mu)}(t)\right]^2}.
 \label{eq:yl-subtracted-Q}
\end{align}
As the non-universal lattice-to-CFT coefficient $A_X$ cancels in this ratio,
we do not need a reference-time normalization. Eq.~\eqref{eq:G_OO=Q_B}
therefore yields the prediction
\begin{align}
 Q_{\widetilde X\widetilde X,\alpha}^{(\mu)}(r,t)
 \simeq
 Q_\mathds{1}\left[x_{\alpha}^{(\mu)}(r,t)\right].
 \label{eq:yl-Q-prediction}
\end{align}
The $\mathcal{S}$- and $\eta$-paired cross ratios $x_\alpha^{(\mu)}(r,t)$ are precisely those defined in Eqs.~\eqref{eq:cross-ratio-S} and~\eqref{eq:eta-cross-ratio}. For a complex $\mathcal{S}$-paired extrapolation parameter, the Euclidean branch of $Q_\mathds{1}(x)$ is analytically continued along the continuous path $x_{\alpha}^{(\mathcal{S})}(r,t)$, as prescribed in Sec.~\ref{sec:general-two-point}.

Comparisons of $Q_{\widetilde X\widetilde X,\alpha}^{(\mu)}$ with
$Q_\mathds{1}(x^{(\mu)}_\alpha)$ are shown in Fig.~\ref{fig:yl-bpz}.
Fig.~\ref{fig:yl-bpz}(a) and (b) show the magnitude and continuously tracked
phase at $r=6$. For product $X$, the two microscopic left covectors coincide;
hence, the $\mathcal{S}$- and $\eta$-paired lattice data are exactly
identical. Because its extrapolation parameter is real, the corresponding
cross ratio stays strictly on the real interval, and $Q_\mathds{1}(x)$
remains real throughout the evolution.

The product-$Z$ and product-$XZ$ preparations, however, distinguish the two geometries. For the $\eta$ pairing, $T_{\alpha}^{(\eta)}$ is real and $x_{\alpha}^{(\eta)}$ remains within the interval $0<x<1$. The resulting ratios are therefore real, and their continuously tracked phases remain zero. For the $\mathcal{S}$ pairing, the imaginary parts of $T_{\alpha}^{(\mathcal{S})}$ drive the cross ratios away from the real axis. The same function $Q_\mathds{1}(x)$ then yields both the modified magnitude and the nonzero phase observed in the lattice data.

Fig.~\ref{fig:yl-bpz}(c) removes the preparation-dependent map from time to the cross ratio by plotting the $\eta$-paired data directly against $x_{\alpha}^{(\eta)}$. Results for $r=4$, $6$, and $8$, across all three preparations, collapse onto the single real function $Q_\mathds{1}(x)$. Thus the preparation and separation dependence of the time-domain
curves is absorbed entirely into the maps
$t\mapsto x_{\alpha}^{(\eta)}(r,t)$; the scaling function itself is
the same $Q_{\mathds{1}}(x)$ for all data sets.

For the $\mathcal{S}$ pairing, the
  common scaling function maps preparation-dependent paths between two complex
  planes. Fig.~\ref{fig:yl-bpz}(d) displays the ratios at $r=6$ as trajectories in the complex $Q$ plane, with arrows indicating increasing time. Product $X$ remains on the real axis, whereas the product-$Z$ and product-$XZ$ preparations follow distinct complex trajectories. The inset shows the associated paths of $x_{\alpha}^{(\mathcal{S})}$ in the cross-ratio plane. Each lattice trajectory is the image of its
preparation-dependent cross-ratio path under the same analytically
continued function $Q_{\mathds{1}}(x)$.

The vertical dashed line in Figs.~\ref{fig:yl-bpz}(a) and~\ref{fig:yl-bpz}(b) marks the conservative
pre-wraparound condition
\begin{align}
 \frac{2vt}{L-r}<0.8.
 \label{eq:yl-prewraparound}
\end{align}
Within this window, the statically subtracted operator gives
percent-level normalized complex RMS residuals for every preparation
and pairing. Appendix~\ref{app:yl-operator-matching} documents the
systematic improvement relative to the bare operator $X_j$.

The spatial correlator therefore tests conformal information not
fixed by the one-point scaling dimension alone. After the local
one-point factors are removed, the remaining dynamics resolve the
boundary-sewn relative block weight and, for a complex extrapolation
parameter, its analytic continuation along the complex cross-ratio
path.

\subsection{Direct field-on quench}
\label{sec:preparation-hierarchy}

The filtered preparations considered above expose the regularized
boundary-state structure directly. We now investigate how the same strip dynamics emerges in a conventional sudden quench. The right preparation $\vert{}\Omega_\lambda\rangle$ is chosen to be the ground state of
\begin{align}
 H_{\rm para}(\lambda)
 =
 -\sum_j\left(Z_jZ_{j+1}+\lambda X_j\right).
 \label{eq:direct-field-initial}
\end{align}
At $t=0$, the critical imaginary longitudinal field
$\mathrm{i}h_z^c(\lambda)$ is switched on, and the state evolves under $H_{\rm YL}(\lambda)$. This realizes a direct field-on quench without any explicit ultraviolet filter.

For $\lambda>1$, this transverse-field Ising Hamiltonian has a unique, real, and parity-even ground state \cite{Pfeuty1970}. Its transpose- and parity-paired covectors therefore coincide, and we suppress the pairing label throughout this subsection. The ground state again flows to the identity Cardy boundary, as do the three product-$\alpha$ preparations.

\begin{figure*}[t]
 \includegraphics[width=\textwidth]
 {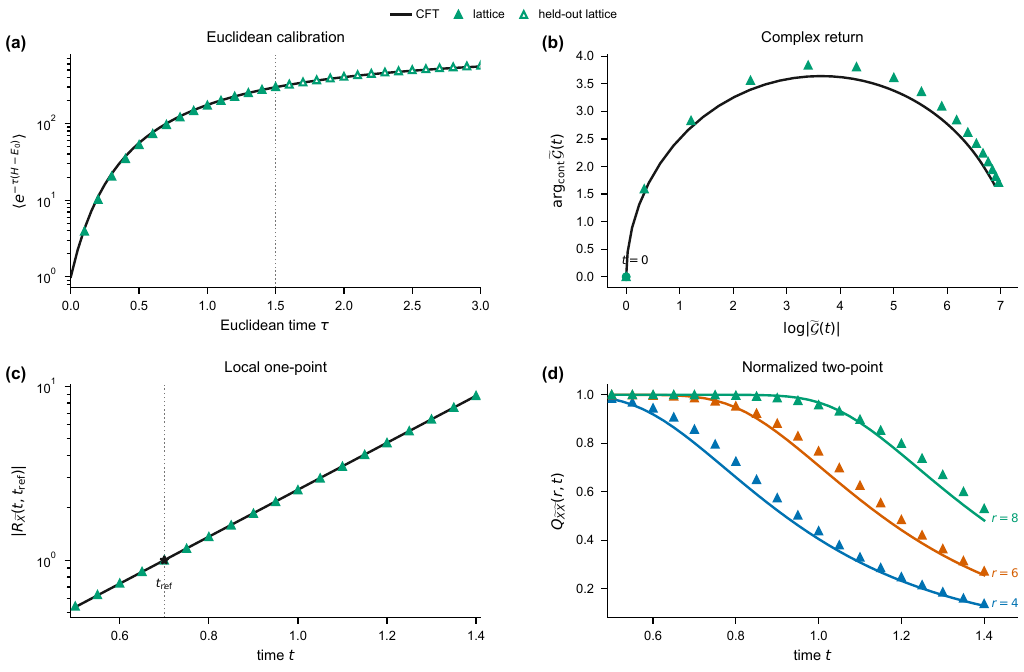}
 \caption{\textbf{Direct field-on Yang--Lee quench at $\lambda=4$.}
 The finite-size ground state defined by
 Eq.~\eqref{eq:direct-field-initial} is quenched directly to
 $H_{\rm YL}(4)$.
 All panels show $L=22$. Solid curves are the BCFT
 predictions and symbols are lattice data.
 (a) Euclidean calibration of $T$, shown with a logarithmic vertical axis.
 Filled triangles to the left of the vertical dotted line enter the
 fit; open triangles are held out.
 (b) The shifted return amplitude represented parametrically by
 $\bigl(\log\vert{}\widetilde{\mathcal G}(t)\vert{},
 \arg_{\rm cont}\widetilde{\mathcal G}(t)\bigr)$, beginning at
 $t=0$.
 (c) Reference-normalized one-point function
 $R_{\widetilde X}(t,t_{\rm ref})$, with $t_{\rm ref}=0.7$ indicated by the
 dotted line and black star. It is real to numerical precision.
 (d) One-point-normalized two-point function
 $Q_{\widetilde X\widetilde X}(r,t)=G_{\widetilde X\widetilde X}(r,t)/
 [m_{\widetilde X}(t)]^2$ for
 $r=4,6,$ and $8$, without any additional time normalization. All
 real-time curves use $T$ from panel (a), the independently determined
 velocity, and the conformal data fixed above.}
 \label{fig:direct-field}
\end{figure*}

The bare ground state retains appreciable ultraviolet corrections at finite size. We therefore determine its real total extrapolation parameter $T$ by fitting the shifted Euclidean amplitude in Eq.~\eqref{eq:app-T-euclidean-amplitude}, using the same $E_0(L)$ shift as in the return amplitude, over a calibration window. Incorporating the full trajectory reduces sensitivity to higher-order finite-size corrections, which can strongly affect a projected-sector ratio evaluated at a single system size. We then freeze $T$ and apply it to every real-time prediction below. The results for $\lambda=4$ are shown in Fig.~\ref{fig:direct-field}.

Fig.~\ref{fig:direct-field}(a) demonstrates that a single real $T$ describes the Euclidean trajectory accurately over the calibration interval and continues quantitatively into the held-out region. The complete trajectory thus provides a stable estimate of the long-distance strip scale.

Freezing this value, Fig.~\ref{fig:direct-field}(b) applies the complete-character return formula in Eq.~\eqref{eq:yl-paired-return}. Although $T$ is real, the return remains complex because real-time continuation moves the cylinder modulus from $T$ to $T+it$. The character prediction captures the amplification, phase winding, and subsequent turning of the lattice trajectory. A systematic displacement remains near the region of maximal phase, indicating that the complete return retains microscopic preparation information beyond a single effective strip scale.

The same Euclidean-calibrated parameter can now be transferred to
observables that did not enter the calibration.
Fig.~\ref{fig:direct-field}(c) applies the midpoint one-point
law in Eq.~\eqref{eq:S-onepoint-ratio} to the local $\widetilde X_j$ using
$x_\phi=-2/5$.  The same $c_{\mathds{1}}(22)$ fixed for the filtered
preparations is used here, because the final Hamiltonian and size are
identical.  Normalization at $t_{\rm ref}$ cancels the remaining
lattice-to-primary coefficient, and the curve tracks the lattice trajectory
over the displayed interval.

Figure~\ref{fig:direct-field}(d) provides a second independent
cross-observable test at the level of
the boundary conformal block. Because $T$ is real, the cross ratio remains
within $0<x<1$, allowing Eq.~\eqref{eq:yl-Q-prediction} to be applied with
$Q_\mathds{1}(x)$ from Eq.~\eqref{eq:yl-Q1}. At a fixed
time, larger separations remain closer to the $x\to1$ limit and hence retain
$Q_{\widetilde X\widetilde X}\simeq1$ for longer. The decline begins first at $r=4$,
followed by $r=6$ and $r=8$, as observed in the lattice data. The identical
block function captures this ordering and the crossover shapes; the normalized complex RMS residuals for $t\ge0.5$ within the
pre-wraparound window of Eq.~\eqref{eq:yl-prewraparound} are $6.01\%$, $4.42\%$, and $2.44\%$ for
$r=4,6,$ and $8$, respectively.

Taken together, Fig.~\ref{fig:direct-field} shows that explicit
Euclidean filtering by the final Hamiltonian is not a prerequisite
for the emergence of conformal quench dynamics. Starting from a
conventional paramagnetic ground state, a single real strip parameter
fixed entirely from Euclidean evolution captures the overall complex
return trajectory and gives quantitative predictions for independent
local and spatial observables, without any real-time adjustment.
The filtering used in the preceding subsections should therefore be
viewed as a controlled means of suppressing ultraviolet corrections,
rather than as an essential ingredient of the temporal-boundary
construction. The appearance of the same strip structure in this
direct field-on quench supports the broader applicability of the
construction beyond explicitly filtered boundary-state preparations.

\section{Complex primary dynamics at the five-state Potts fixed point}
\label{sec:potts}

The non-Hermitian five-state Potts chain provides a cross-model test in which
complexity enters both ingredients of the one-point law~\eqref{eq:general-onepoint-law}: the temporal
strip parameter and the scaling dimension of the inserted operator.
For real couplings, the ferromagnetic five-state Potts transition is
weakly first order, while its nearby scaling behavior is organized by
a conjugate pair of fixed points at complex coupling. Continuing the
self-dual quantum chain to one member of this pair gives a
non-Hermitian critical Hamiltonian described by a complex CFT, with
complex scaling dimensions and boundary data
\cite{GorbenkoRychkovZan2018PartI,GorbenkoRychkovZan2018,Nienhuis1984,TangEtAl2024,TangBoundary,VanderLindenEtAl2026,LiuEtAl2026}.

Here, our objective is more focused than in the Yang--Lee benchmark. We use static projected data to fix a single complex temporal geometry and then ask whether it correctly predicts the magnitude and phase evolution of two distinct primary fields. Static spectra and real-time dynamics are evaluated at the largest accessible size, $L=10$, using exact diagonalization and matrix-exponential action in
the zero-momentum sector.

\subsection{Complex fixed point and projected temporal geometry}

With clock operators
\begin{align}
 \sigma\vert{}n\rangle
 =
 \mathrm{e}^{2\pi\mathrm{i}n/5}\vert{}n\rangle,
 \qquad
 \tau\vert{}n\rangle
 =
 \vert{}n+1\!\!\!\pmod 5\rangle,
\end{align}
the periodic Hamiltonian reads
\begin{align}
    H_{\rm Potts}
  =
  H_0+\lambda H_1,
\end{align}
with
\begin{align}
    H_0
  &=
  -
  \sum_{j=1}^{L}
  \sum_{m=1}^{4}
  \left[
    \left(\sigma_j^\dagger \sigma_{j+1}\right)^m
    +
    \tau_j^m
  \right],\notag\\
  H_1
  &=
  \sum_{j=1}^{L}
  \sum_{m,n=1}^{4}
  \Bigl[
    \left(\tau_j^m+\tau_{j+1}^m\right)
    \left(\sigma_j^\dagger \sigma_{j+1}\right)^n\notag\\
    &+
    \left(\sigma_j^\dagger \sigma_{j+1}\right)^m
    \left(\tau_j^n+\tau_{j+1}^n\right)
  \Bigr]
\end{align}
with periodic boundary condition~\cite{Wu1982,TangEtAl2024}.
At the complex critical point~\cite{VanderLindenEtAl2026}
\begin{align}
 \lambda_c\approx0.0788+0.0603\,\mathrm{i},
\end{align}
the resulting Hamiltonian is complex symmetric,
\begin{align}
 H_{\rm Potts}^{\mathsf T}
 =
 H_{\rm Potts}
 \neq
 H_{\rm Potts}^{\dagger}.
\end{align}
We therefore use the $\mathcal S$ pairing with
$\mathcal S=\mathds{1}$.

Starting from the fixed-color product state
\begin{align}
 \vert{}B_{\rm fixed}\rangle=\vert{}0\rangle^{\otimes L},
\end{align}
we prepare
\begin{align}
 \vert{}R_{\rm fixed}\rangle
 &:=
 \mathrm{e}^{-\frac{\beta}{2}
 [H_{\rm Potts}-E_0(L)]}
 \vert{}B_{\rm fixed}\rangle,
 \nonumber\\
 \langle L_{\rm fixed}^{(\mathcal S)}\vert{}
 &:=
 {\vert{}R_{\rm fixed}\rangle}^{\mathsf T},
 \qquad
 \beta=1,
 \label{eq:potts-filtered-preparation}
\end{align}
where $E_0(L)$ is the finite-size level flowing to the CFT
identity and a Euclidean filter is added as in the Yang--Lee product states.

We determine the total extrapolation parameter of the filtered
fixed-color preparation using the basis-invariant projected-sector
construction summarized in Appendix~\ref{app:T-projected}. Previous
boundary spectroscopy identifies this preparation with the fixed
Cardy boundary, whose coefficient ratio is known from analytic
continuation of the $Q<4$ Potts boundary data
\cite{TangBoundary,LiuEtAl2026}. Using the identity sector and the
complete four-dimensional spin multiplet, we obtain at $L=10$
\begin{align}
 T_{\rm fixed}^{(\mathcal S)}(L=10)
 =
 1.298266+0.261890\,\mathrm{i}.
 \label{eq:potts-width-ten}
\end{align}
This is the total extrapolation parameter of the filtered
preparation and already includes the unit Euclidean filter. We
suppress the size argument below and hold this value fixed in all
real-time comparisons.

\subsection{Two primary fields on the same complex strip}

We combine the total extrapolation parameter in Eq.~\eqref{eq:potts-width-ten} with the known
complex scaling dimensions~\cite{GorbenkoRychkovZan2018,TangEtAl2024}
\begin{align}
 x_\sigma
 &\approx
 0.133597-0.020464\,\mathrm{i},\notag\\
 x_\epsilon
 &\approx
 0.465613-0.224495\,\mathrm{i}.
 \label{eq:potts-dimensions}
\end{align}
The spin field is represented by
\begin{align}
 O_\sigma(j)=\sigma_j,
\end{align}
while a lattice representative of the energy field is
\begin{align}
 O_\epsilon(j)
 =
 \sum_{k=1}^{4}
 \left[
 \left(\sigma_j^\dagger\sigma_{j+1}\right)^k
 -
 \tau_j^k
 \right].
 \label{eq:potts-energy}
\end{align}
No identity subtraction is required for these Potts observables:
$O_\sigma$ carries nontrivial $\mathbb Z_5$ charge, while the difference in
Eq.~\eqref{eq:potts-energy} is odd under self-duality at the self-dual point;
neither can mix with the duality-even, neutral identity operator
\cite{Wu1982,TangEtAl2024}.

\begin{figure*}[t]
 \includegraphics[width=\textwidth]
 {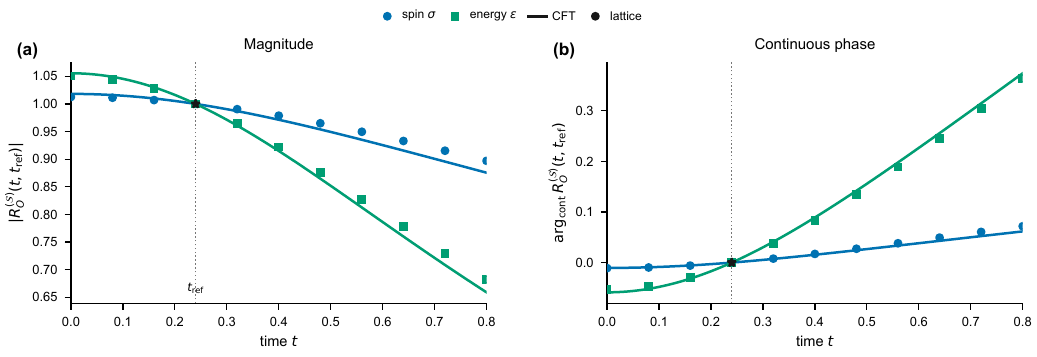}
 \caption{\textbf{Complex primary dynamics at the five-state Potts
 fixed point.}
 Magnitudes (a) and continuously tracked phases (b) of the
 reference-normalized spin and energy one-point functions for the
 filtered fixed-color preparation at $\beta=1$ and $L=10$.
 Blue circles and green squares denote the spin and energy lattice
 data, respectively, and solid curves are the CFT predictions.
 The vertical dotted line marks $t_{\rm ref}=0.24$, where both
 normalized ratios have unit magnitude and zero phase. All curves use the same complex extrapolation parameter fixed from
the static projected spin multiplet, together with the independently
known complex scaling dimensions.}
 \label{fig:potts}
\end{figure*}

For $\mathcal{O}=\sigma,\epsilon$, the transpose-paired midpoint law
in Eq.~\eqref{eq:S-onepoint-ratio} gives
\begin{align}
 R_O^{(\mathcal S)}(t,t_{\rm ref})
 &:={}
 \frac{\langle O(t)\rangle}{\langle O(t_{\rm ref})\rangle}
 \notag\\
 &\simeq
 \left[
 \frac{
 \cosh\!\left(\pi t/T_{\rm fixed}^{(\mathcal S)}\right)
 }{
 \cosh\!\left(\pi t_{\rm ref}/T_{\rm fixed}^{(\mathcal S)}\right)
 }
 \right]^{-x_\mathcal{O}}.
 \label{eq:potts-onepoint-prediction}
\end{align}
Here we take $t_{\rm ref}=0.24$.
The logarithm is transported continuously from
$t=t_{\rm ref}$, where
$R_O^{(\mathcal S)}(t_{\rm ref},t_{\rm ref})=1$.
Thus the same complex function controls both
observables, while their different complex dimensions determine how
that geometry is resolved into magnitude and phase.

The common normalization at $t=t_{\rm ref}$ is fixed by definition;
the nontrivial comparison is the simultaneous magnitude and phase
evolution away from that point. Fig.~\ref{fig:potts}(a) shows a
gentle decrease of the spin response and a substantially stronger
decrease of the energy response. Fig.~\ref{fig:potts}(b) shows the corresponding phase evolution.
The spin phase remains close to zero throughout the displayed
interval, whereas the energy response accumulates a substantially
larger continuous phase. Since both $T_{\rm fixed}^{(\mathcal S)}$ and $x_O$ are complex,
neither component is controlled by a single part of the scaling
dimension: $\operatorname{Re}x_O$ and $\operatorname{Im}x_O$
contribute jointly to both the magnitude and the phase.

Using the single parameter $T$ fixed before real-time evolution, the CFT
curves reproduce the curvature and relative separation of both
operator trajectories. The continuous phases are especially well
captured, while the lattice magnitudes develop a modest smooth upward
displacement from the leading-strip curves at the latest displayed
times. Since $T$ was extracted from the static spin multiplet,
the energy trajectory constitutes an independent cross-sector
prediction.

This comparison tests the direct transfer
from static boundary data to dynamics for a non-Hermitian model whose universal class is represented as a complex CFT.
Unlike the rational Yang--Lee minimal model, however, the
fixed-boundary Potts cylinder amplitude does not truncate to the
finite Virasoro-character combination used above, and the accessible
size leaves too narrow a space--time window for a controlled
two-point test. We therefore use the Potts calculation as a focused
test of complex primary dynamics.

Nevertheless, the fact that a parameter calibrated in the spin
sector also predicts the energy trajectory shows that the
local-primary construction is not tied to the rational Yang--Lee
theory or to a single operator sector. It supports a broader
applicability to non-Hermitian critical systems, including complex
CFTs with complex scaling dimensions, whenever the relevant
temporal-boundary data can be determined independently.

\section{Discussion and outlook}
\label{sec:discussion}

The central result of this work is a BCFT organization of biorthogonal global quenches in genuinely interacting non-Hermitian critical systems. In the language of the boundary renormalization group, the microscopic left and right preparations flow to macroscopic temporal conformal boundaries, with their generally complex one-sided extrapolation parameters encoding the leading displacement from the boundary fixed point. Their real parts govern the Euclidean regularization familiar from Hermitian quenches, whereas their imaginary parts explicitly retain coherent microscopic preparation phases. Depending on the precise microscopic symmetry pairing, this identical phase manifests as a complex strip modulus, a displacement of the real-time insertion, or a specified kinematic trajectory through the conformal moduli space. Negative scaling dimensions permit exponential growth of primary
one-point functions in the infinite-strip regime. For an
$\eta$-paired preparation, the same biorthogonal expectation can
also be written as a ratio of Dirac-normalized expectations:
\begin{align}
 \langle O(t)\rangle_{\eta}
 =
 \frac{
 \langle \eta O\rangle_{RR,t}
 }{
 \langle\eta\rangle_{RR,t}
 },
 \qquad
 \langle \cdot\rangle_{RR,t}
 :=
 \frac{
 \langle R(t)|\cdot|R(t)\rangle
 }{
 \langle R(t)|R(t)\rangle
 }.
 \label{eq:discussion-eta-RR-map}
\end{align}
The biorthogonal representation keeps the conformal insertion local,
whereas Eq.~\eqref{eq:discussion-eta-RR-map} expresses the same
expectation value as a ratio of Dirac-normalized expectations of the
symmetry-dressed operators $\eta O$ and $\eta$, which are generally
nonlocal. Related trade-offs between locality and unitarity arise in
similarity transformations between local non-Hermitian and nonlocal
Hermitian descriptions
\cite{Mostafazadeh2010,CastroAlvaredoFring2009}, and in mappings between
nonunitary and unitary critical theories
\cite{GuruswamyLudwig1998,HsiehChang2023,FukusumiKawamoto2025}.

The temporal-boundary picture suggests two direct extensions and
several broader applications.  If the left and right preparations
flow to distinct boundary conditions, $B_L\neq B_R$, the cylinder
amplitude becomes $Z_{B_LB_R}$, and the corresponding mixed boundary
sector---equivalently, boundary-condition-changing operators in the
upper-half-plane representation---enters the correlators and sewing
relations.  At finite size, the leading strip formulas should be
replaced by annulus correlators with complex modulus set by $vT/L$.
This extension would carry the local and spatial predictions beyond
the pre-wraparound regime and may clarify the relation between
return-amplitude zeros, revivals, and dynamical quantum phase
transitions.  The same organization may also be useful in other
non-Hermitian critical theories and in inhomogeneous quenches with
spatially dependent one-sided parameters.

Entanglement dynamics constitute a distinct yet fundamentally related frontier. Evaluating R{\'e}nyi or pseudoentropy dynamics requires replicating the microscopic left--right pairing together with twist-field or defect insertions at the entanglement cut. Because these quantities are not strictly fixed by the single-sheet character and correlator data established here, constructing and solving these replicated temporal boundaries stands as a compelling next step.

Taken together, our results identify the infrared left and right
boundary conditions, the complex one-sided extrapolation parameters,
and the static lattice-to-field matching as the data that organize
the non-Hermitian critical quenches studied here.  Formulated directly
in BCFT rather than through a free-particle decomposition, this
framework connects phase-sensitive microscopic preparations to
universal real-time dynamics and provides a basis for extending the
construction to broader classes of interacting non-Hermitian critical
systems.

\begin{acknowledgments}
Y.~L. thanks Kohei Kawabata and Masaki Oshikawa for helpful discussions and comments on the manuscript. Y.~L. is supported by the Global Science Graduate Course (GSGC) program of the University of Tokyo. This work was supported by JSPS KAKENHI Grant Number JP23K25791.
\end{acknowledgments}

\appendix

\section{Static calibrations and numerical conventions}
\label{app:T}
This appendix details the static calibrations and numerical conventions used in the main text. The central step in our setup is the determination of the total extrapolation parameter,
\begin{align}
    T=\tau_L+\tau_R,
\end{align}
which governs the cylinder amplitude for a fixed pair of left and right preparations. For $\mathcal{S}$-paired preparations used in the main text, the relations in Eq.~\eqref{eq:paired-one-sided-parameters} uniquely fix the one-sided geometry once $T^{(\mathcal{S})}$ is known. For independent left and right preparations, the two one-sided parameters must instead be calibrated separately, as illustrated in Appendix~\ref{app:mixed-preparations}.

To determine this total parameter prior to real-time evolution, we introduce two different approaches: one based on projected spectral weights, which isolate the propagation between selected conformal sectors, and another based on a Euclidean-time trajectory, which utilizes the complete cylinder partition function. Both methods rely strictly on finite-size spectral information, but differ in their requisite BCFT data: the former requires information from only two selected sectors, whereas the latter relies on the complete cylinder partition function.

Following the explanation of these calibrations and their model-specific implementations, we summarize the numerical evolution and analytic-continuation procedures, and document the static matching of the Yang--Lee lattice operator.

\subsection{Projected-sector determination}
\label{app:T-projected}

For the final critical non-Hermitian Hamiltonian considered in this work, let a finite-size eigenspace $i$ be spanned by right and left eigenvectors
$|\Psi^R_{i,a}\rangle$ and $\langle\Psi^L_{i,a}|$, where $a$ denotes the degeneracy. Its overlap matrix and
basis-invariant spectral projector are defined as
\begin{align}
 S^{(i)}_{ab}
 &:=
 \langle\Psi^L_{i,a}|\Psi^R_{i,b}\rangle,
 \\
 \Pi_i
 &:=
 \sum_{a,b}
 |\Psi^R_{i,a}\rangle
 \bigl[(S^{(i)})^{-1}\bigr]_{ab}
 \langle\Psi^L_{i,b}| .
 \label{eq:app-T-projector}
\end{align}
The overlap matrix $S^{(i)}$ encodes the biorthogonal normalization,
and $\Pi_i$ is the projector onto eigenspace $i$. The weight of the
preparation in this eigenspace is
\begin{align}
 Z_i^{\rm lat}
 &:=
 \langle L_0|\Pi_i|R_0\rangle .
 \label{eq:app-T-sector-weight}
\end{align}
This definition is invariant under any eigenvector normalization and
nonsingular basis change within a degenerate eigenspace. As the preparation is fixed, we omit this dependence in Eq.~\eqref{eq:app-T-sector-weight}. For a
nondegenerate energy level $i$, it reduces to the product of the two preparation overlaps
divided by the normalization $\langle\Psi_i^L|\Psi_i^R\rangle$.

For the boundary condition selected by the preparation, write the
closed-channel cylinder amplitude as
\begin{align}
 Z_B(T)
 =
 \sum_i G_i\,\chi_i[q(T)],
 \qquad
 q(T)=\exp\left(-\frac{4\pi v T}{L}\right),
 \label{eq:app-T-cylinder}
\end{align}
where the signed or complex coefficients $G_i$ are fixed BCFT data.
Keeping the finite-size propagation of the primary representatives in two
sectors gives
\begin{align}
 \frac{Z_i^{\rm lat}}{Z_j^{\rm lat}}
 \simeq
 \frac{G_i}{G_j}
 \exp\left\{-T_{ij}(L)
 [E_i(L)-E_j(L)]\right\}.
 \label{eq:app-T-sector-ratio}
\end{align}
The corresponding finite-size estimate is therefore
\begin{align}
 T_{ij}(L)
 =
 -\frac{1}{E_i(L)-E_j(L)}
 \operatorname{Log}_{\rm cont}
 \left[
 \frac{Z_i^{\rm lat}G_j}
      {Z_j^{\rm lat}G_i}
 \right].
 \label{eq:app-T-projected}
\end{align}
This coefficient ratio must retain its BCFT sign or phase factor. The initial
logarithmic image is fixed by a stated static branch criterion; at each
subsequent point of a size or coupling sequence, the image closest to the
preceding value in the complex $T$ plane is retained. The numerical branch
choices used below are specified explicitly. 
Different sector pairs give the same $T$ in the scaling limit and provide a direct finite-size consistency check when available. This method requires only the eigenvalues and eigenvectors of two
sectors, $i$ and $j$, together with the BCFT coefficient ratio
$G_i/G_j$. It is therefore practical for irrational theories, where
only partial CFT data may be available from bootstrap or analytic
continuation, as in the non-Hermitian five-state Potts example.

\subsection{Euclidean-trajectory determination}
\label{app:T-euclidean}

When the complete cylinder partition function is available, $T$ can instead be
determined without isolating individual sectors. Choose a reference energy
$E_{\rm ref}(L)$ and evaluate the normalized Euclidean amplitude
\begin{align}
 \mathcal E_L(\tau)
 &:=
 \frac{
 \langle L_0|
 e^{-\tau[H-E_{\rm ref}(L)]}
 |R_0\rangle
 }{\langle L_0|R_0\rangle},
 \\
 \mathcal E_{\rm CFT}(\tau;T)
 &:=
 e^{\tau E_{\rm ref}^{\rm CFT}(L)}\frac{Z_B(T+\tau)}
 {Z_B(T)},
 \label{eq:app-T-euclidean-amplitude}
\end{align}
where $Z_B$ denotes the cylinder partition function and $E_{\rm ref}^{\rm CFT}(L)$ denotes the CFT counterpart
of the lattice reference-energy shift $E_{\rm ref}(L)$. When
the shift is chosen as the energy of a finite-size state flowing
to scaling dimension $x_{\rm ref}$, its CFT value is
\begin{align}
 E_{\rm ref}^{\rm CFT}(L)
 =
 \frac{2\pi v}{L}
 \left(
 x_{\rm ref}-\frac{c}{12}
 \right).
\end{align}

For a calibration set $\mathcal C$ of Euclidean times, we may determine the strip parameter through
\begin{align}
 T(L)
 =
 \operatorname*{arg\,min}_{T\in\mathcal D}
 \frac{
 \sum_{\tau\in\mathcal C}
 |\mathcal E_L(\tau)-\mathcal E_{\rm CFT}(\tau;T)|^2
 }{
 \sum_{\tau\in\mathcal C}|\mathcal E_L(\tau)|^2
 } .
 \label{eq:app-T-euclidean-fit}
\end{align}
The domain $\mathcal D$ enforces $\operatorname{Re}T>0$ and any reality
condition implied by the preparation. When $T$ is complex, the analytically continued CFT amplitude is
evaluated on a branch transported continuously from the specified
initial calibration point. By incorporating all contributing conformal towers and multiple Euclidean times, this method reduces the sensitivity to sector-dependent finite-size corrections in a single projected ratio, at the cost of requiring the complete cylinder partition function and performing an imaginary-time evolution. Euclidean times outside $\mathcal C$ provide a
held-out check, as in Fig.~\ref{fig:direct-field} (a).

\subsection{Model-specific calibrations used in the main text}
\label{app:T-numerics}

\paragraph{Filtered Yang--Lee preparations.}
For each even size $L=12,14,\ldots,22$, we project independently with the
$\mathcal S$- and $\eta$-paired preparations onto the nondegenerate $\phi$ and identity
primary sectors. Periodic-chain spectroscopy fixes
$G_{\mathds{1}}/G_\phi=-\varphi$ with
$\varphi=(1+\sqrt{5})/2$ \cite{LiuEtAl2026}. Eq.~\eqref{eq:app-T-projected}
is then evaluated with $(i,j)=(\mathds{1},\phi)$ and the same-size energy
gap. At $L=12$ we choose the branch with the smallest
$|\operatorname{Im}T|$; the branch at each larger even size is selected by
continuity with the preceding value.
The symmetric filter acts for $\beta/2$ on each one-sided extrapolation parameter, so applying
the projection after filtering is equivalently (and also exactly) described by
$T_\beta=T_0+\beta$. The two pairings are processed independently; their
agreement with Eq.~\eqref{eq:yl-T-relation} is therefore not imposed in the
extraction. The values extracted at $L=22$ and quoted in
Sec.~\ref{sec:yl-extrapolation-parameters} are used in
every $L=22$ Yang--Lee real-time prediction throughout the main text.

\paragraph{Direct field-on preparation.}
For the $\lambda=4$, $L=22$ quench in Fig.~\ref{fig:direct-field}, the bare
paramagnetic ground state has a real total parameter. We use the complete
signed Yang--Lee character combination, the spectroscopic velocity, and the
same $E_0(L)$ shift as in Eqs.~\eqref{eq:yl-paired-return} and
\eqref{eq:yl-return-character}. The single positive
parameter in Eq.~\eqref{eq:app-T-euclidean-fit} is calibrated over
$0.05\leq\tau\leq1.50$; the disjoint interval
$1.55\leq\tau\leq3.00$ is held out. This gives $T\approx 0.4043$, which is then
frozen in the return, one-point, and two-point comparisons.

\paragraph{Five-state Potts preparation.}
At $L=10$, the identity representative is nondegenerate, whereas the
spin sector is a four-dimensional multiplet due to the $S_5$ symmetry. We retain the complete
multiplet in Eq.~\eqref{eq:app-T-projector} and combine
$Z_\sigma^{\rm lat}/Z_{\mathds{1}}^{\rm lat}$ with the fixed-color
boundary-coefficient ratio obtained in
Ref.~\cite{LiuEtAl2026}. Continuous tracking of the projected phase
along the available size sequence selects logarithmic branch $n=0$.
A direct comparison of all logarithmic images at $L=10$ selects the
same branch and gives the intrinsic value
\begin{align}
 T_0^{(\mathcal S)}
 \approx
 0.298266+0.261890\,\mathrm{i}.
\end{align}
The symmetric $\beta=1$ filter adds one real unit, yielding the total
parameter quoted in Eq.~\eqref{eq:potts-width-ten}.

\subsection{Numerical implementation and analytic-continuation
conventions}
\label{app:numerics}

All lattice calculations reported here exploit translation
invariance and are performed in the zero-momentum orbit basis.
Within the zero-momentum sector, translation invariance makes the matrix elements of $O_j$ and
$O_jO_{j+r}$ independent of the chosen site $j$. The required low-energy
right and left eigenspaces are obtained with
\texttt{scipy.sparse.linalg.eigs}
\cite{LehoucqSorensenYang1998,VirtanenEtAl2020}. Spectral weights for degenerate sectors use the
basis-invariant projector in Eq.~\eqref{eq:app-T-projector}.
Eigensystem residuals, biorthogonal overlaps, and the conditioning of
the overlap matrix are checked before inversion. The largest
calculations use $L=22$ for the Yang--Lee chain and $L=10$ for the
five-state Potts chain.

Euclidean filters and real-time states are evaluated through sparse
matrix-exponential action using
\texttt{scipy.sparse.linalg.expm\_multiply}, without explicitly
forming the matrix exponential
\cite{AlMohyHigham2011,VirtanenEtAl2020}. For the Yang--Lee
calculations, the implementation uses
\begin{align}
 H_s
 &=
 \frac{H_{\rm YL}-E_0(L)}{\Delta_L},
 t_s
 =
 \Delta_L t,
 \nonumber\\
 \Delta_L
 &=
 E_1(L)-E_0(L),
\end{align}
so that $H_st_s=[H_{\rm YL}-E_0(L)]t$. The five-state Potts
calculations act with the unshifted Hamiltonian $H_{\rm Potts}$.
A scalar energy shift cancels from normalized equal-time one- and
two-point functions, while it remains in the shifted Yang--Lee return
and Euclidean amplitude, so a shifted comparison~\eqref{eq:general-return-strip} is employed in the main text.

The Yang--Lee trajectories are sampled on a uniform grid with
$\Delta t=0.025$, while the Potts trajectories use $\Delta t=0.01$;
these values are output sampling intervals. The Yang--Lee actions are
checked by independent evaluations at
overlapping intermediate times. The Potts evolution is checked
through conservation of the left--right overlap. The corresponding
relative consistency errors are below $10^{-8}$. We separately verify
\begin{align}
 \widetilde{\mathcal G}(0)=1,
 \qquad
 R_O(t_{\rm ref},t_{\rm ref})=1.
\end{align}

For a complex trajectory $Y(t)$ sampled at times $t_n$ in an interval
$\mathcal I$, we define the normalized complex RMS residual by
\begin{align}
 \epsilon_Y(\mathcal I)
 &:=
 \left[
 \frac{
 \sum_{t_n\in\mathcal I}
 \left|
 Y_{\rm lat}(t_n)-Y_{\rm CFT}(t_n)
 \right|^2
 }{
 \sum_{t_n\in\mathcal I}
 \left|
 Y_{\rm lat}(t_n)
 \right|^2
 }
 \right]^{1/2}.
 \label{eq:app-complex-rms}
\end{align}
The same sampling times are used whenever two calibrations or
operator definitions are compared.

The generic branch-transport rule for the projected logarithm is
given below Eq.~\eqref{eq:app-T-projected}, while the model-specific
initial branches are specified in
Sec.~\ref{app:T-numerics}. Return phases are unwrapped continuously
and anchored to zero at $t=0$. Complex powers in the one-point law
are evaluated through a logarithm transported continuously from
$t=t_{\rm ref}$. The Yang--Lee hypergeometric blocks are continued
from the Euclidean interval along the ordered complex cross-ratio
path.

\subsection{Static matching of the Yang--Lee lattice operator}
\label{app:yl-operator-matching}

The bare lattice operator $X_j$ contains both identity and Yang--Lee primary
components.  We isolate the latter through the static subtraction in
Eq.~\eqref{eq:yl-identity-subtracted-X}.  The coefficient
$c_{\mathds{1}}(L)$ is obtained solely from an identity-sector matrix element
of the final Hamiltonian and is independent of the quench preparation,
pairing, and real-time data.  Independent evaluations using the biorthogonal left and right
eigenvectors, the complex-symmetric transpose pairing, and the
$\eta=P$ pairing give the same coefficient within the numerical
precision reported below.

For a normalization-independent measure of the identity component,
we define
\begin{align}
 \rho_{\mathds{1}/\phi}(L)
 &:=
 \frac{|c_{\mathds{1}}(L)|}{A_{\phi}^{\rm gi}(L)},
 \nonumber\\
 A_{\phi}^{\rm gi}(L)
 &:=
 \left|
 \frac{
 \langle\Psi_\phi^L|X_j|\Psi_{\mathds{1}}^R\rangle
 \langle\Psi_{\mathds{1}}^L|X_j|\Psi_\phi^R\rangle
 }{
 \langle\Psi_\phi^L|\Psi_\phi^R\rangle
 \langle\Psi_{\mathds{1}}^L|\Psi_{\mathds{1}}^R\rangle
 }
 \right|^{1/2}.
 \label{eq:app-yl-identity-primary-ratio}
\end{align}
Table~\ref{tab:app-yl-operator-matching} shows that
$c_{\mathds{1}}(L)$ is numerically modest but increases slightly over
the available sizes, whereas the gauge-invariant identity-to-primary
ratio decreases from $4.10\%$ to $3.54\%$.

\begin{table*}[!t]
 \caption{Static Yang--Lee operator matching.  The matrix
 element defining $c_{\mathds{1}}(L)$ is given in
 Eq.~\eqref{eq:yl-identity-subtracted-X}, and
 $\rho_{\mathds{1}/\phi}$ is defined in
 Eq.~\eqref{eq:app-yl-identity-primary-ratio}.}
 \label{tab:app-yl-operator-matching}
 \begin{ruledtabular}
 \begin{tabular}{rccc}
 $L$ & $\operatorname{Re}c_{\mathds{1}}$
 & $\operatorname{Im}c_{\mathds{1}}$
 & $\rho_{\mathds{1}/\phi}$ \\
 \hline
 12 & $0.05241009$ & $3.69\times10^{-13}$ & $4.101\%$ \\
 14 & $0.05443680$ & $3.87\times10^{-12}$ & $4.008\%$ \\
 16 & $0.05569450$ & $8.23\times10^{-12}$ & $3.890\%$ \\
 18 & $0.05652191$ & $2.40\times10^{-11}$ & $3.768\%$ \\
 20 & $0.05709314$ & $3.67\times10^{-10}$ & $3.650\%$ \\
 22 & $0.05750468$ & $-6.59\times10^{-11}$ & $3.540\%$ \\
 \end{tabular}
 \end{ruledtabular}
\end{table*}

Existing one- and two-point data are converted algebraically as
\begin{align}
 m_{\widetilde X}(t)
 &=m_X(t)-c_{\mathds{1}}(L),
 \nonumber\\
 G_{\widetilde X\widetilde X}(r,t)
 &=G_{XX}(r,t)-2c_{\mathds{1}}(L)m_X(t)
 +c_{\mathds{1}}(L)^2.
 \label{eq:app-yl-operator-postprocessing}
\end{align}
As shown in
Tab.~\ref{tab:app-yl-operator-residuals}, the same static subtraction
systematically improves every preparation and pairing. Here and below, residuals are evaluated using the normalized complex
RMS metric defined in Eq.~\eqref{eq:app-complex-rms}.

\begin{table*}[!t]
 \caption{Normalized complex RMS residuals at $L=22$ before
 and after static operator matching.  One-point residuals use
 $0.5\leq t\leq1.4$; two-point residuals are evaluated at $r=6$ within the pre-wraparound window defined in Eq.~\eqref{eq:yl-prewraparound}, with the additional restriction $t\geq 0.5$ to exclude the initial constant region. For product $X$, the $\mathcal{S}$- and $\eta$-paired data are identical.}
 \label{tab:app-yl-operator-residuals}
 \setlength{\tabcolsep}{4pt}
 \begin{ruledtabular}
 \begin{tabular}{llcc}
 Observable & preparation/pairing & $O=X$ & $O=\widetilde X$ \\
 \hline
 $R_O$ & $X$, $\mathcal S=\eta$ & $1.185\%$ & $0.227\%$ \\
 $R_O$ & $Z$, $\mathcal S$ & $1.833\%$ & $0.981\%$ \\
 $R_O$ & $Z$, $\eta$ & $2.665\%$ & $1.172\%$ \\
 $R_O$ & $XZ$, $\mathcal S$ & $1.373\%$ & $0.431\%$ \\
 $R_O$ & $XZ$, $\eta$ & $1.676\%$ & $0.468\%$ \\
 \hline
 $Q_{OO}$ & $X$, $\mathcal S=\eta$ & $0.942\%$ & $0.630\%$ \\
 $Q_{OO}$ & $Z$, $\mathcal S$ & $1.792\%$ & $1.489\%$ \\
 $Q_{OO}$ & $Z$, $\eta$ & $1.415\%$ & $1.175\%$ \\
 $Q_{OO}$ & $XZ$, $\mathcal S$ & $1.170\%$ & $0.868\%$ \\
 $Q_{OO}$ & $XZ$, $\eta$ & $0.917\%$ & $0.681\%$ \\
 \end{tabular}
 \end{ruledtabular}
\end{table*}

At $r=4,6,$ and $8$, the largest pre-wraparound changes induced by
the subtraction in $|Q_{XX}|$ are $0.0106$, $0.00714$, and $0.00231$,
respectively. The larger effect at shorter separation is consistent
with ultraviolet operator mixing. We therefore use
$\widetilde X_j(L)$ throughout the Yang--Lee comparisons rather than
treating the identity component of the bare operator as negligible, keeping in mind that the scaling of the original one- or two-point functions can be recovered through Eq.~\eqref{eq:app-yl-operator-postprocessing}
explicitly.
\section{Finite-size flow of the Yang--Lee temporal-boundary
calibration}
\label{app:finite-size-T}

The projected ratio in Eq.~\eqref{eq:app-T-projected}
defines an effective finite-size temporal-boundary parameter
$T(L)$. Here we analyze its finite-size flow for the
filtered Yang--Lee preparations. As specified in
Sec.~\ref{app:T-numerics}, all $L=22$ main-text curves use
$T_\alpha^{(\mu)}(22)$ and $v(22)$ obtained at that same size;
$T_{\alpha,\infty}^{(\mu)}$ is used here only to quantify the
finite-size drift and assess its asymptotic range.

\begin{figure}[!t]
 \includegraphics[width=\columnwidth]
 {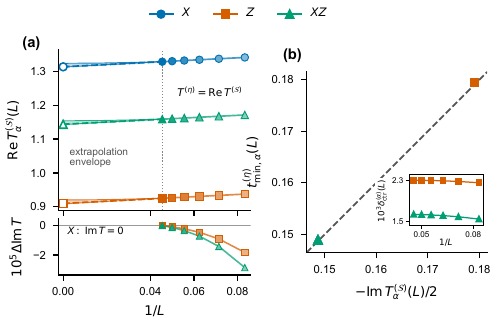}
 \caption{\textbf{Finite-size flow of the Yang--Lee temporal-boundary
 calibration.}
 (a) Filled symbols show
 $\operatorname{Re}T_\alpha^{(\mathcal S)}(L)$ for the three filtered
  preparations at $L=12,14,\ldots,22$, and solid segments are guides to the
  eye. The vertical dotted line marks $1/L=1/22$; the region to its left is
  extrapolative. There, the colored bands span the predictions of all
  retained fits in Eq.~\eqref{eq:app-T-finite-size-ansatz}, the dashed curves
  give their pointwise median, and the open symbols at $1/L=0$ denote the
  median intercepts. The lower subpanel resolves
 $10^5\Delta\operatorname{Im}T_\alpha^{(\mathcal S)}(L)$, where
 $\Delta\operatorname{Im}T_\alpha^{(\mathcal S)}(L)
 =\operatorname{Im}T_\alpha^{(\mathcal S)}(L)
 -\operatorname{Im}T_\alpha^{(\mathcal S)}(22)$; for product $X$ this
 quantity vanishes identically.
 (b) Independently observed minima of the $\eta$-paired
identity-subtracted one-point magnitude
 plotted against the static prediction
 $-\operatorname{Im}T_\alpha^{(\mathcal S)}(L)/2$ for product $Z$ and the
 $XZ$ preparation. The dashed line denotes equality. Markers at different sizes largely overlap
 within the symbol width in the main panel. The inset resolves
 $10^3\delta_{\rm ctr}^{(\alpha)}(L)$ as a function of $1/L$.}
 \label{fig:app-yanglee-finite-size}
\end{figure}

We estimate its asymptotic range by fitting the real and imaginary
components separately to
\begin{align}
 T_\alpha^{(\mu)}(L)
 &=T_{\alpha,\infty}^{(\mu)}+\frac{a_\alpha^{(\mu)}}{L},
 \nonumber\\
 T_\alpha^{(\mu)}(L)
 &=T_{\alpha,\infty}^{(\mu)}+\frac{a_\alpha^{(\mu)}}{L^2},
 \nonumber\\
 T_\alpha^{(\mu)}(L)
 &=T_{\alpha,\infty}^{(\mu)}+\frac{a_\alpha^{(\mu)}}{L}
   +\frac{b_\alpha^{(\mu)}}{L^2}.
 \label{eq:app-T-finite-size-ansatz}
\end{align}
The one-correction forms use the final four, five, and six sizes, while the
two-correction form uses the final five and six. Because the projected-ratio
construction does not fix the leading correction exponent, the spread over
these choices only defines a systematic extrapolation envelope rather than a
statistical confidence interval. For the $\mathcal S$ pairing, it gives
\begin{align}
 1.313\lesssim\operatorname{Re}T_{X,\infty}^{(\mathcal S)}
 &\lesssim1.324,\notag
 \\
 0.909\lesssim\operatorname{Re}T_{Z,\infty}^{(\mathcal S)}
 &\lesssim0.920,\notag
 \\
 1.143\lesssim\operatorname{Re}T_{XZ,\infty}^{(\mathcal S)}
 &\lesssim1.154.\notag
\end{align}
The corresponding imaginary parts remain near $-0.3581$ and $-0.2974$ for
$Z$ and $XZ$, with a spread below $7\times10^{-5}$. The independently
extracted $\eta$ sequence obeys Eq.~\eqref{eq:yl-T-relation} to numerical
precision. Fig.~\ref{fig:app-yanglee-finite-size} summarizes these
finite-size flows and the independent dynamical test discussed below.

Finite-$L$ comparisons retain the size-matched calibration. Replacing only
$T_\alpha^{(\mu)}(22)$ by an asymptotic intercept while retaining the
$L=22$ spectrum, matrix elements, velocity, and dynamics mixes ingredients
from different finite-size descriptions. To quantify this effect, we use one
retained member of the extrapolation family, the six-size two-correction fit
\begin{align}
 T_\alpha^{(\mu)}(L)
 &=T_{\alpha,\infty}^{(\mu)}
 +\frac{a_\alpha^{(\mu)}}{L}
 +\frac{b_\alpha^{(\mu)}}{L^2},
 \qquad L=12,14,\ldots,22.
 \label{eq:app-T-illustrative-fit}
\end{align}
This choice enters only the stability comparison below and no main-text
curve. We use the $Z$, $\mathcal S$-paired observables as a representative
test because their nonzero imaginary temporal parameter makes the return
phase and temporal displacement particularly sensitive to this substitution.

Residuals in
Tab.~\ref{tab:app-extrapolated-T-comparison}
are evaluated using the normalized complex RMS metric in
Eq.~\eqref{eq:app-complex-rms}. The return comparison uses
$0\leq t\leq1.4$. For the local observables, we use the common interval
$0.5\leq t\leq1.4$. The two-point values therefore need not coincide with
those in Tab.~\ref{tab:app-yl-operator-residuals}, which use the full
pre-wraparound window. The unmatched substitution preserves the qualitative
winding, phase ordering, and conformal-block trajectory, while the residuals
in Tab.~\ref{tab:app-extrapolated-T-comparison} support retaining the
size-matched calibration in the finite-$L$ main-text comparison.
\begin{table*}[!t]
 \caption{Normalized complex RMS residuals for $Z$, $\mathcal S$-paired
 observables at $L=22$, as defined in Eq.~\eqref{eq:app-complex-rms}. The
 extrapolated column uses Eq.~\eqref{eq:app-T-illustrative-fit}; all other
 finite-size inputs are unchanged.}
 \label{tab:app-extrapolated-T-comparison}
 \begin{ruledtabular}
 \begin{tabular}{lcc}
 Observable and interval
 & $T_Z^{(\mathcal S)}(22)$
 & $T_{Z,\infty}^{(\mathcal S)}$ \\
 \hline
 Return, $0\leq t\leq1.4$ & $4.78\%$ & $11.77\%$ \\
 $R_{\widetilde X}$, $0.5\leq t\leq1.4$ & $0.98\%$ & $1.87\%$ \\
 $Q_{\widetilde X\widetilde X}(r=6)$, $0.5\leq t\leq1.4$ & $1.38\%$ & $1.73\%$ \\
 \end{tabular}
 \end{ruledtabular}
\end{table*}

The temporal-center relation supplies an independent check along the same
finite-size flow, without introducing a real-time fitting parameter. For each
$L=12,14,\ldots,22$, we use the corresponding static
$c_{\mathds{1}}(L)$ fixed in Appendix~\ref{app:yl-operator-matching} and
locate the minimum of $|R_{\widetilde X,\alpha}^{(\eta)}|$ by a local
seven-point quartic interpolation. For
$\alpha\in\{Z,XZ\}$,
\begin{align}
 t_{\rm min,\alpha}^{(\eta)}(L)
 &\simeq
 -\frac{1}{2}\operatorname{Im}T_\alpha^{(\mathcal S)}(L),
 \label{eq:app-T-finite-size-center}
 \\
 \delta_{\rm ctr}^{(\alpha)}(L)
 &:=
 \frac{
 \left|
 t_{\rm min,\alpha}^{(\eta)}(L)
 +\frac{1}{2}\operatorname{Im}T_\alpha^{(\mathcal S)}(L)
 \right|
 }{
 \left|\operatorname{Im}T_\alpha^{(\mathcal S)}(L)\right|/2
 }.
 \label{eq:app-center-deviation}
\end{align}
$\delta_{\rm ctr}^{(\alpha)}(L)$ remains below $2.32\times10^{-3}$; for $X$
both sides of Eq.~\eqref{eq:app-T-finite-size-center} vanish identically.

Thus the imaginary component of the temporal-boundary parameter and
its transfer to an independently defined dynamical time center remain
stable throughout the finite-size flow, even though the extrapolated
real-part intercept retains a modest dependence on the fitting form and
window. This separation motivates the size-matched parameters used in the
main-text finite-$L$ comparisons.

\section{Mixed left and right preparations}
\label{app:mixed-preparations}

The strip construction in Sec.~\ref{sec:framework} does not require
the microscopic left and right preparations to be related by either
of the symmetry pairings in Eq.~\eqref{eq:yl-paired-covectors}. It
applies equally to independent preparations, provided that the two
preparations flow to the same infrared conformal boundary condition.
We test this generality by combining the filtered Yang--Lee
preparations in Eq.~\eqref{eq:yl-filtered-preparation}, all of which
flow to the identity Cardy boundary. A label
$\alpha_L\mid\alpha_R$ is ordered from the left covector to the right
ket. Unless stated otherwise, all results use
$L=22$, $\lambda=4$, and Euclidean filter $\beta=1$ as in the main text.

For each preparation $\alpha\in\{X,Z,XZ\}$, the
$\mathcal S$-paired projected parameter determines the one-sided extrapolation
parameter $\tau_R$,
\begin{align}
 \tau_{R,\alpha}
 &=
 \frac12T_\alpha^{(\mathcal S)},
 \nonumber\\
 \tau_{L,\alpha}^{(\mathcal S)}
 &=
 \tau_{R,\alpha},
 \qquad
 \tau_{L,\alpha}^{(\eta)}
 =
 \tau_{R,\alpha}^{*}.
 \label{eq:app-mixed-leg-parameters}
\end{align}
For a mixed pair
$\alpha_L^{(\mu_L)}\mid\alpha_R$, the strip geometry is therefore
assembled as
\begin{align}
 T_{\alpha_L^{(\mu_L)}\mid\alpha_R}
 &:=
 \tau_{L,\alpha_L}^{(\mu_L)}
 +
 \tau_{R,\alpha_R},
 \nonumber\\
 u_{\alpha_L^{(\mu_L)}\mid\alpha_R}(t)
 &:=
 \tau_{R,\alpha_R}+\mathrm{i}t.
 \label{eq:app-mixed-geometry}
\end{align}
No superscript is shown for product $X$ as the two left
covectors coincide. The quantities in
Eq.~\eqref{eq:app-mixed-geometry} are inserted directly into
Eqs.~\eqref{eq:general-return-strip},
\eqref{eq:general-onepoint-ratio}, and
\eqref{eq:G_OO=Q_B}.

For the first four pairs in
Tab.~\ref{tab:app-mixed-preparations}, we also determine the total
parameter directly by applying Eq.~\eqref{eq:app-T-projected} to the
mixed left covector and right ket. The directly projected value and
the value assembled from the two independently calibrated extrapolation parameters agree
to a relative difference below $4\times10^{-10}$, which works as an algebraic and branch-consistency check of the one-sided calibration.

For the final two rows, we additionally construct a right ket by
taking the Hermitian conjugate of the $\eta$-paired covector,
\begin{align}
 |R_Z^{(\eta\dagger)}\rangle
 &:=
 \left(
 \langle L_Z^{(\eta)}|
 \right)^\dagger
 =
 P|R_Z\rangle,
 \nonumber\\
 |R_{Z,\mathrm f}^{(\eta\dagger)}\rangle
 &:=
 \mathrm{e}^{-\frac{\beta_{\rm add}}{2}
 [H_{\rm YL}-E_0(L)]}
 |R_Z^{(\eta\dagger)}\rangle,
 \qquad
 \beta_{\rm add}=1.
 \label{eq:app-mixed-dagger-right}
\end{align}
The first identity is exact microscopically, but it does not determine
the right-sided extrapolation parameter. The relations in
Eq.~\eqref{eq:paired-one-sided-parameters} follow because the
$\mathcal S$ and $\eta$ maps intertwine the Hamiltonians defining the
two extrapolation parameters. Hermitian conjugation, by contrast, turns a
preparation generated by $H_{\rm YL}$ into one generated by
$H_{\rm YL}^{\dagger}$. When the resulting ket is subsequently used
as a right preparation for evolution with $H_{\rm YL}$, no symmetry
relation fixes its $\tau_R$, which must therefore be calibrated
independently.

We first determine the total parameter
$T_{X\mid Z^{(\eta\dagger)}}$ from the static projected ratio using
the product-$X$ left covector. Since its left-sided parameter
$\tau_{L,X}$ is already known, this fixes
\begin{align}
 \tau_{R,Z}^{(\eta\dagger)}
 =
 T_{X\mid Z^{(\eta\dagger)}}-\tau_{L,X}.
 \label{eq:app-mixed-dagger-leg}
\end{align}
The resulting right-sided parameter is a property of the right
preparation and is transferred unchanged to the
$Z^{(\mathcal S)}$ left covector. Without the additional filter, we
obtain
\begin{align}
 \tau_{R,Z}^{(\eta\dagger)}
 &\approx
 0.072317+0.017726\,\mathrm{i},
 \nonumber\\
 T_{Z^{(\mathcal S)}\mid Z^{(\eta\dagger)}}
 &\approx
 0.534869-0.161309\,\mathrm{i}.
 \label{eq:app-mixed-dagger-parameters}
\end{align}
The additional one-sided filter in
Eq.~\eqref{eq:app-mixed-dagger-right} increases
$\operatorname{Re}\tau_R$ by
$\beta_{\rm add}/2=0.5$ while leaving its imaginary part unchanged.
It therefore shifts the total parameter to
\begin{align}
 T_{Z^{(\mathcal S)}\mid Z_{\mathrm f}^{(\eta\dagger)}}
 \approx
 1.034869-0.161309\,\mathrm{i}.
 \label{eq:app-mixed-filtered-dagger-T}
\end{align}

\begin{table*}[!t]
 \caption{\textbf{Mixed-preparation results at $L=22$.}
 Pair labels are ordered from the left covector to the right ket, and
 $T$ is fixed entirely by static projected data. The last three
 columns give normalized complex RMS residuals as defined in
 Eq.~\eqref{eq:app-complex-rms}; the two-point column uses
 $Q_{\widetilde X\widetilde X}(r=6,t)$.}
 \label{tab:app-mixed-preparations}
 \small
 \setlength{\tabcolsep}{3.5pt}
 \begin{ruledtabular}
 \begin{tabular}{lcccc}
 Pair
 &
 $T$
 &
 $\widetilde{\mathcal G}$
 &
 $R_{\widetilde X}$
 &
 $Q_{\widetilde X\widetilde X}(r=6)$
 \\
 \hline
 $X\mid Z$
 &
 $1.126877-0.179035\,\mathrm{i}$
 &
 $2.77\%$
 &
 $0.49\%$
 &
 $0.24\%$
 \\
 $X\mid XZ$
 &
 $1.243752-0.148718\,\mathrm{i}$
 &
 $1.93\%$
 &
 $0.32\%$
 &
 $0.18\%$
 \\
 $XZ^{(\mathcal S)}\mid Z$
 &
 $1.041979-0.327753\,\mathrm{i}$
 &
 $3.45\%$
 &
 $0.65\%$
 &
 $0.37\%$
 \\
 $XZ^{(\eta)}\mid Z$
 &
 $1.041979-0.030316\,\mathrm{i}$
 &
 $3.60\%$
 &
 $0.72\%$
 &
 $0.17\%$
 \\
 $Z^{(\mathcal S)}\mid Z^{(\eta\dagger)}$
 &
 $0.534869-0.161309\,\mathrm{i}$
 &
 $25.21\%$
 &
 $5.60\%$
 &
 $6.12\%$
 \\
 $Z^{(\mathcal S)}\mid Z_{\mathrm f}^{(\eta\dagger)}$
 &
 $1.034869-0.161309\,\mathrm{i}$
 &
 $3.68\%$
 &
 $0.67\%$
 &
 $0.62\%$
 \\
 \end{tabular}
 \end{ruledtabular}
\end{table*}

The return residuals in
Tab.~\ref{tab:app-mixed-preparations} use
$0\leq t\leq1.4$, while the $R_{\widetilde X}$ residuals use
$0.5\leq t\leq1.4$ with $t_{\rm ref}=0.7$. For
$Q_{\widetilde X\widetilde X}(r=6,t)$, we use the
pre-wraparound window
\begin{align}
 \frac{2vt}{L-r}\leq0.6.
\end{align}
The same statically determined $c_{\mathds{1}}(22)$ is used for every
mixed pair.

\begin{figure*}[t]
 \includegraphics[width=\textwidth]
 {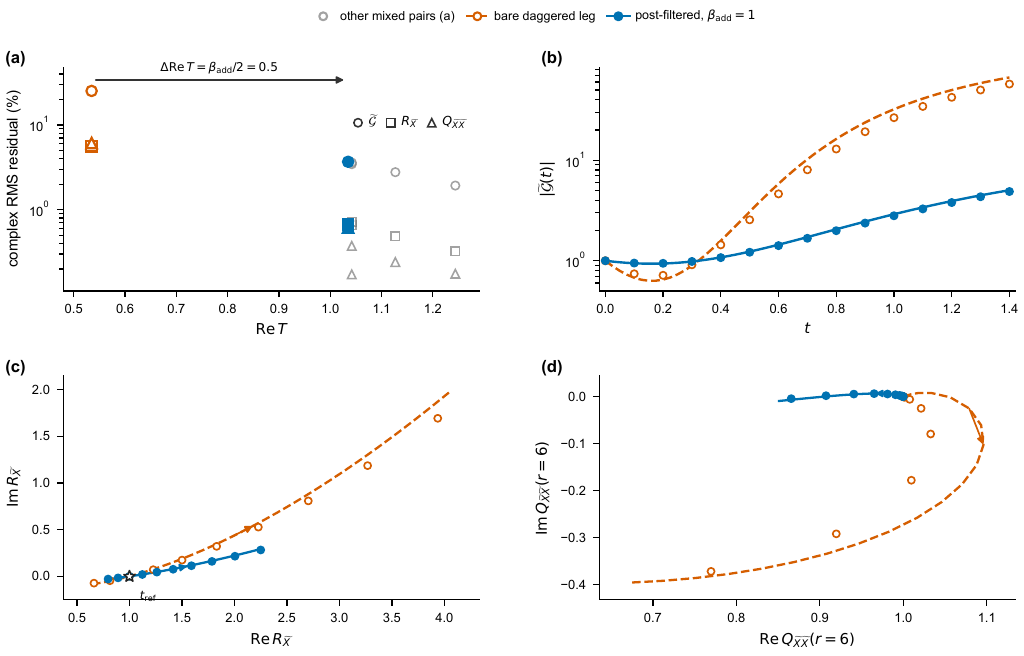}
 \caption{\textbf{Mixed-preparation dynamics and control of the
 Euclidean width.}
 (a) Normalized complex RMS residuals plotted against
 $\operatorname{Re}T$. Gray symbols denote the first four rows of
 Tab.~\ref{tab:app-mixed-preparations}; orange open and blue filled
 symbols denote the bare and post-filtered daggered right preparations in the
 fifth and sixth rows. Circles, squares, and triangles represent
 $\widetilde{\mathcal G}$, $R_{\widetilde X}$, and
 $Q_{\widetilde X\widetilde X}$, respectively. The arrow shows the
 fixed shift
 $\Delta\operatorname{Re}T=\beta_{\rm add}/2=0.5$ produced by the
 additional one-sided filter.
 (b) Return magnitudes for
 $Z^{(\mathcal S)}\mid Z^{(\eta\dagger)}$ before and after the
 additional filter.
 (c) Corresponding complex
 $R_{\widetilde X}(t,t_{\rm ref})$ trajectories; the star marks
 $t_{\rm ref}=0.7$.
 (d) Complex
 $Q_{\widetilde X\widetilde X}(r=6,t)$ trajectories over the common
 conservative window $2vt/(L-r)\leq0.60$.
 In panels (b)--(d), lines are the strip-CFT predictions and symbols
 are lattice data. Dashed lines with open symbols denote the bare
 daggered right preparation, while solid lines with filled symbols denote the
 post-filtered preparation. Arrows indicate increasing time. All
 temporal-boundary parameters are fixed by static projected data.}
 \label{fig:app-yanglee-mixed-preparations}
\end{figure*}

The first four rows in Tab.~\ref{tab:app-mixed-preparations} show that the strip predictions remain
quantitative when the two microscopic preparations are chosen
independently. The bare daggered right preparation in the fifth row instead
has a small real strip width and correspondingly large
finite-size and ultraviolet corrections. The additional one-sided
filter increases only the real part of the calibrated geometry and
simultaneously reduces the return, one-point, and two-point residuals
in the sixth row. This controlled comparison supports the short
Euclidean width, rather than the use of independent left and
right preparations, as the dominant source of the preceding
deviations. Fig.~\ref{fig:app-yanglee-mixed-preparations} resolves this
comparison across the global and local dynamical observables.

Together, these results show that the symmetry pairings are a
parameter-reducing specialization rather than a prerequisite of the
strip construction. Once the two one-sided extrapolation parameters
are calibrated independently, the same return, one-point, and
two-point formulas remain predictive for independent microscopic
preparations that flow to the same conformal boundary. The one-sided
filtering control further separates this generality from ultraviolet
corrections associated with a small Euclidean strip width.

\bibliography{references}

\end{document}